\documentclass[fleqn,usenatbib]{mnras}
\usepackage{newtxtext,newtxmath}
\usepackage{float} 
\usepackage[T1]{fontenc}
\usepackage{ae,aecompl}
\usepackage{natbib}

\usepackage{graphicx}	
\usepackage{amsmath}	
\usepackage{amssymb}	

\usepackage{array,multirow}
\usepackage{comment}
\usepackage{xcolor}
\definecolor{lava}{rgb}{0.81, 0.06, 0.13}

\title[Non-thermal emission from BH-disc impacts]{
A hadronic emission model for black hole-disc impacts in the blazar OJ 287
}

\author[J. C. Rodr\'iguez-Ram\'irez et al.]{
J. C. Rodr\'iguez-Ram\'irez,$^{1}$\thanks{E-mail: juan.rodriguez@iag.usp.br}
P. Kushwaha,$^{1,2}$\thanks{E-mail:pankaj.tifr@gmail.com}
E. M. de Gouveia Dal Pino$^{1}$\thanks{E-mail: 
dalpino@iag.usp.br }
and
\newauthor
R. Santos-Lima$^{1}$\thanks{E-mail: lima.reinaldo.santos.de@gmail.com }
\\
$^{1}$Instituto de Astronomia,  Geof\'isica e Ci\^{e}ncias Atmosf\'ericas (IAG-USP),
             Universidade de S\~{a}o Paulo.
             S\~ao Paulo-SP 05508-090, Brasil.\\
$^{2}$ Aryabhatta Research Institute of Observational Sciences (ARIES). Nainital 263002, India.
}

\date{Accepted 2020 August 26. Received 2020 August 23; in original form 2020 May 1.}

\pubyear{2020}

\begin{document}
\label{firstpage}
\pagerange{\pageref{firstpage}--\pageref{lastpage}}
\maketitle

\begin{abstract}
A super-massive black hole (SMBH) binary in the core of the blazar OJ 287 has been invoked in
previous works to explain its observed optical flare quasi-periodicity.
Following this picture, we investigate a hadronic origin for the X-ray and $\gamma$-ray counterparts of the November 2015 major optical flare of this source. 
An impact outflow must result after the lighter SMBH (the secondary) crosses the 
accretion disc of the heavier one (the primary).
We then consider acceleration of cosmic-ray (CR) protons in the 
shock driven by the impact outflow as it expands and collides with the active galactic nucleus 
(AGN) wind of the primary SMBH. 
We show that the emission of these CRs can reproduce the X-ray and  $\gamma$-ray 
flare data self-consistently with the optical component of the November 2015 major flare.
The derived emission models are
consistent with a magnetic field $B \sim 5$ G in the emission region 
and a power-law index of $q\sim2.2$
for the energy distribution of the emitting CRs.
The mechanical luminosity of the AGN wind 
represents $\lesssim 50\%$ of the mass accretion power
of the primary SMBH in all the derived emission profiles.
\end{abstract}




\section{Introduction}
\label{sec:intro}
Theoretical arguments as well as indirect observational evidence suggest the presence of
super-massive black hole (SMBH) pairs coalescing in the core of certain galaxies.
Galaxy mergers \citep{2005Natur.435..629S}, for instance, might be a natural process leading to the formation of such SMBH binaries.
Compelling  examples of active galactic nuclei (AGNs)
approaching each other can be found in the recent works by,
e. g., \cite{2019ApJ...883..167P} and \cite{2014Natur.511...57D}, where the SMBHs of  approaching AGNs are localised at distances 
 from tens to hundreds of parsecs between each other.

 When the distance among two SMBHs shrinks to sub-parsec scales,  the system is theoretically expected to enter
 its gravitational wave (GW)-driven regime for orbital decay.
 In such a stage, SMBH binaries are thought to be the most prominent sources of GWs in the cosmos 
\citep{1980Natur.287..307B, 2017NatAs...1..886M}.
Current instruments however, are not able to detect either GWs from SMBHs systems (expected in the nHz-$\mu$Hz domain), or resolve SMBHs binaries at sub-parsec scales.
Alternatively, indirect signatures as double line emission \citep{2012NewAR..56...74P} and  quasi-periodical flares 
 in certain AGNs \citep{2016IAUS..312...13K} are employed to trace the presence of compact, orbiting SMBH pairs.
Due to a persistent quasi-periodical feature in optical, the blazar OJ 287 is perhaps the strongest candidate for hosting a sub-parsec SMBH pair \citep{2018ApJ...866...11D}.

OJ 287 (at a red-shift $z = 0.306$) is categorised as a BL Lac object and is known for its regular $\sim$12 year, double peaked optical variations registered for over 130 years \citep{1988ApJ...325..628S, 2013A&A...559A..20H}.
These periodic features have motivated a number of possible explanations
\citep[e.g.,][]{1996ApJ...460..207L, 1997ApJ...478..527K, 2013MNRAS.434.2275T, 2018MNRAS.478.3199B}.
Particularly, the SMBH binary scenario  proposed by 
\cite{1996ApJ...460..207L} \citep[see also][]{2008Natur.452..851V}
appears to predict naturally the timing of the double peaked observed outbursts.
Additionally, this model is consistent with the sharp rise of the flare emission and its low polarisation degree, being these aspects not satisfactorily explained by other models
\citep[see][for more details]{2019Univ....5..108D, 2020Galax...8...15K}.

The SMBH binary model of \cite{1996ApJ...460..207L} explains the
periodical outbursts of OJ 287 in terms of thermal bremsstrahlung
radiation of the outflows generated by the impacts of the lighter SMBH
(the secondary) on the accretion disc of the heavier one 
\citep[the primary, see also][]{2016MNRAS.457.1145P}. 
Within this picture, a general relativistic (GR) approach for the orbit
of the secondary BH predicted the starting times of the 
1994, 1995, 2005, 2007, and 2015 flares \citep{2008Natur.452..851V,2016ApJ...819L..37V}. 
With the observed data from the last three outbursts, the BH masses of
the the binary have been constrained to 
$M_1=1.83\times 10^{10}$ M$_\odot$ and $M_2 = 1.5\times 10^{8}$ 
M$_\odot$ for the primary and secondary BHs, respectively \citep{2016ApJ...819L..37V}.

While the analysis developed in,  
\cite{1996ApJ...460..207L},
\cite{1998ApJ...507..131I}, and
\cite{2016MNRAS.457.1145P}
is applicable to the problem of a BH threading quasi-perpendicularly 
an accretion disc, other studies in the context of star-disc collisions
(and their observational consequences) 
can be found in, e. g.,
\cite{1983Ap&SS..95...11Z},
\cite{2004A&A...413..173N}, and
\cite{2016ApJ...823..155K}.

As expected from BL Lac objects, OJ 287 displays X-ray as well as $\gamma$-ray flaring behaviour
\citep{2011MNRAS.412.1389N, 2017A&A...597A..80H, 2013MNRAS.433.2380K, 2018MNRAS.473.1145K, 2018MNRAS.479.1672K, 2020ApJ...890...47P}.
Particularly, \cite{2018MNRAS.473.1145K} analyse the multi-wavelength (MW)
light curves (LCs) of OJ 287 during and after the November 2015 major optical flare. 
These authors  extracted the corresponding spectral energy distribution (SED) of the flare and interpreted it with 
a leptonic, jet emission model. They found that the  X-ray component of the flare is well explained by synchrotron self-Compton (SSC) emission, whereas the $\gamma$-ray flare component is better explained with external Compton (EC) emission  \citep[see also][]{2018bhcb.confE..22K}.

In the present paper we consider the observed MW SED obtained by \cite{2018MNRAS.473.1145K} and 
alternatively investigate a hadronic origin for the high energy (HE) counterpart
(the simultaneous X-ray and $\gamma$-ray excess) of the 
November 2015 optical flare.
We follow the SMBH-disc impact model of 
\cite{1996ApJ...460..207L} \citep[see also][]{2016MNRAS.457.1145P}
and explain the X-ray and $\gamma$-ray fluxes
with emission triggered by proton-proton ($p$-$p$) interactions of cosmic-rays (CRs) with the thermal ions within the impact outflow. 
In the scenario proposed here, we consider CR shock acceleration driven by
the collision of the outflow and the AGN wind of the primary SMBH, as depicted in Figure \ref{fig:out-shock}.

This paper is organised as follows. In the next section, we characterise the SMBH-disc impact outflow and its thermal radiation 
following the considerations of previous works.
In Section 3, we describe the non-thermal radiation that results due to $p$-$p$ interactions of CRs with the thermal ions of the outflow.
In Section 4, we apply the non-thermal emission model to explain the MW
SED corresponding to the 2015 major flare of OJ 287.
We finally summarise and discuss our results in Section 5.

\section{The outflow from the SMBH-disc impact}
\label{sec:bh-disc}
\subsection{The outflow thermal flare}
\label{sec:scenario}

\begin{figure}
	\includegraphics[width=8.5cm]{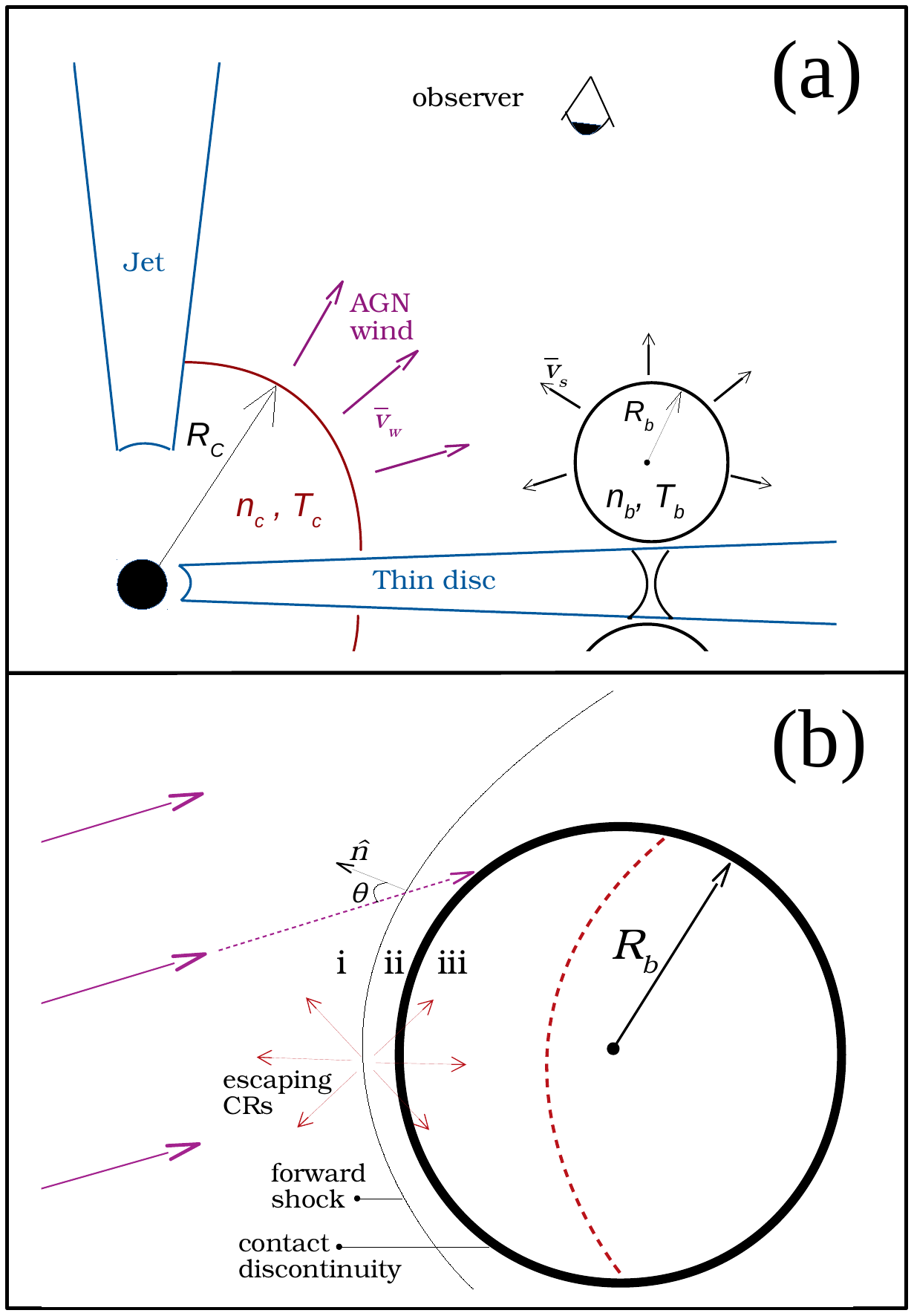}
    \caption{
    (a) Schematic illustration of an idealised spherical outflow that expands above the accretion disc of the primary SMBH.
    The outflow is thought to be generated by a previous impact of the secondary SMBH on the accretion disc of the primary (see the text).
    In the scenario proposed here, the impact outflow forms a shock due to its interaction 
    with the AGN wind driven by the primary SMBH, whose poloidal components are represented with the purple arrows.
    (b) Amplified view of the spherical outflow of the upper panel.     
    The thick solid circle 
    represents the boundary of the outflow material and the thin solid curve is the driven shock.
    The dashed red curve
    represents the boundary of the CR emission volume proposed in this work (see the text).
    The regions of this schematic figure are: 
    (i) the unperturbed AGN wind of density $n_\text{w}$, 
    (ii) the swept up shell of the shocked wind material, and
    (iii) the CR emission volume within the outflow bubble of density $n_\text{b}$.
        }
    \label{fig:out-shock}
\end{figure}

After the secondary SMBH threads the accretion disc of the primary,
two outflows emerge, one above and the other bellow the accretion disc
at the location of the impact.
This effect was simulated by \cite{1998ApJ...507..131I} with a  
hydrodynamical approach. 
Following  \cite{1998ApJ...507..131I} and \cite{2016MNRAS.457.1145P}, 
here we assume a BH-disc impact event that produces a bipolar outflow.
Additionally, we consider that only the outflow emerging in the side of the disc 
pointing toward us contributes to the observed outburst SED.
The analysis described in the following is centred on this
``observer side'' outflow  which we model as a spherical expanding bubble.
A cartoon of this outflow is depicted in
Figure \ref{fig:out-shock}a. 
Within this picture, after the BH-disc impact occurs, a bubble
emerges from the disc with an initial radius $R_0$, gas density
$\rho_0$, and temperature $T_0$.

Here we parametrise the initial radius of the bubble as a fraction $f_\text{R}$
of the the Bondi-Hoyle-Lyttleton radius $R_\text{HL}$
\citep{1939PCPS...35..405H,2004NewAR..48..843E} 
of the secondary SMBH of mass $M_2$, i.e:
\begin{equation}
\label{R_0}
R_0 = f_\text{R} R_\text{HL};\,\, f_\text{R} \leq 1,
\end{equation}
with
\begin{equation}
\label{R_HL}
    R_\text{HL} = \frac{2 G M_2}{v_\text{r}^2},
\end{equation}
where
$v_\text{r}$ is the velocity of the disc material in the co-moving frame of the travelling
BH at the impact event\footnote{Considering the toroidal component of the disc velocity $\Vec{v}_\phi$ and the 
velocity of the secondary SMBH $\vec{v}_\text{orb}$ 
at the location of the impact,
the velocity of the disc material in the co-moving  frame of the secondary BH is
$v_\text{r} \sim |\vec{v}_\text{orb} - \vec{v}_\phi|$ (see Appendix~\ref{appendix:E_outflow}).
}.
Following \cite{1996ApJ...460..207L}, we estimate the initial temperature and density
of the outflow bubble with the jump conditions for a strong,
radiation dominated shock (\citealt{1966_Pai};
where this shock is driven by the secondary BH
during its passage through the accretion disc, see Appendix~\ref{appendix:E_outflow}):
\begin{equation}
\label{T_0}
T_0 = \left(\frac{18 \rho_\text{d} v_\text{r}^2}{7a}\right)^{1/4},
\end{equation}
\begin{equation}
\label{rho_0}
\rho_0 = \left( \frac{\gamma_\text{a} +1}{\gamma_\text{a}-1} \right) \rho_\text{d} =  7\rho_\text{d}.
\end{equation}
In equations (\ref{T_0})-(\ref{rho_0}) $\rho_\text{d}$ is the gas density of the disc at
the location of the impact, 
$a$ is the radiation constant, and
 $\gamma_\text{a}= 4/3$ is the adiabatic index appropriate 
for a radiation-dominated mixture. 
The initial energy of the emerging outflow
can be estimated as
(see Appendix~\ref{appendix:E_outflow} for details):
\begin{align}
\nonumber
    E_0 = &\frac{14\pi}{3} \left(f_\text{R} R_\text{HL}\right)^3 \rho_\text{d} v_\text{r}^2
    = 8.3 \times 10^{55} \times \\
\label{E_0}
    & \left( \frac{M_2}{1.5\times 10^{8} \mbox{M}_\odot}\right)^3
    \left( \frac{f_\text{R}}{0.6}\right)^{3}
   \left( \frac{n_\text{d}}{10^{14}\,\mbox{cm}^{-3}}\right)
    \left( \frac{0.15 c}{v_\text{r}}\right)^{4}
    \mbox{erg},    
\end{align}
where the quantities in the second equality are
normalised with typical values for the parameters of the claimed 
binary system in OJ 287 \citep{2019ApJ...882...88V}.

In an adiabatic expansion, when the bubble attains a radius $R_\text{b}>R_0$ 
its temperature $T_\text{b}$ and gas density $\rho_\text{b}$ are:
\begin{align}
\label{T_b}
    T_\text{b} &= T_0 \xi^{3(1-\gamma_\text{a})},\\
\label{rho_b}
    \rho_\text{b} &= \rho_0 \xi^{-3},
\end{align}
where $\xi\equiv R_\text{b} / \ R_0$.

Similarly to \cite{1996ApJ...460..207L}(see also \citealt{2019ApJ...882...88V}),
we consider that the optical outburst is produced after the spherical bubble
expands adiabatically and the effective optical depth $\tau_\text{e}$ meets the transparency condition:
\begin{equation}
\label{tau_e}
\tau_\text{e} = \sqrt{\kappa_\text{a}\kappa_\text{T}} \, \rho_\text{b} R_\text{b} = 1.
\end{equation}
In equation (\ref{tau_e}), 
 $\kappa_\text{T}$
is the electron opacity and
$\kappa_\text{a}$ is the
frequency-averaged opacity due to absorption.
For a fully ionised gas $\kappa_\text{T} = \sigma_\text{T}/(\mu_\text{e} m_\text{u})$ 
where $\sigma_\text{T}$ is the electron scattering cross section,
$m_\text{u}$ is the atomic mass constant,  
$\mu_\text{e}=2/(1+X)$, and $X$ 
the hydrogen mass fraction. 
According to  \cite{1996ApJ...460..207L} (see also \citealt{2019ApJ...882...88V}),
the absorption opacity $\kappa_\text{a}$ within the outflow bubble 
follows Kramer's law with contributions due to free-free and 
bound-free opacities derived from Rosseland mean.
In this approach, the absorption opacity can be written as:
\begin{equation}
\label{kappa_a}
\kappa_\text{a} = K_\text{a} \rho_\text{b} T_\text{b}^{-7/2}\,\,\text{cm}^2 \text{g}^{-1}, 
\end{equation}
with
\begin{equation}
\label{K_a}
    K_\text{a} =  3.7 \times 10^{22} (1+X) 
    \left[
    1 + \left(\frac{1180}{t_\text{bf}} -1\right)Z 
    \right],
\end{equation}
where $X$ and $Z$ are the mass fractions of hydrogen and metals, respectively, 
and $t_\text{bf}$ is known as the guillotine factor (correction for quantum effects 
of bound-free transitions)
which takes values between 100 and 1 \citep{2007adc..book.....I}.

Condition (\ref{tau_e}) together  with equations (\ref{T_b})-(\ref{rho_b}) and 
(\ref{kappa_a})-(\ref{K_a})
define the radius of the expanding outflow at which the thermal flare is produced:
\begin{equation}
    \label{cond_thin}
   \frac{R_\text{b}}{R_0}\equiv \xi = \left(K_\text{a} \kappa_\text{T} R_0^2 \rho_0^3 T_0^{-7/2}\right)^{2/7},
\end{equation}
which is determined by the initial radius $R_0$, density $\rho_0$, and temperature $T_0$ of the outflow.

Given the values of 
$v_\text{r}$, $\rho_\text{d}$, and $f_\text{R}$, one can determine the
properties of the outflow bubble when it produce the outburst.
First, $R_0$, $\rho_0$, and $T_0$ are obtained with equations (\ref{R_0})-(\ref{rho_0}),
and then $R_\text{b}$, $T_\text{b}$, and $\rho_\text{b}$ through equations 
(\ref{cond_thin}), (\ref{T_b}), and (\ref{rho_b}).
The appropriate combination of the parameters
$v_\text{r}$, $\rho_\text{d}$, and $f_\text{R}$ can be constrained
by fitting the implied thermal bremsstrahlung emission of the bubble of
radius $R_\text{b}$, density $\rho_\text{b}$, and temperature $T_\text{b}$ 
to the observed SED data of the optical flare (see Section \ref{sec:SED_models}).
To do this, we calculate the observed flux due to thermal bremsstrahlung radiation as
\begin{equation}
\label{nuFnu}
    \nu F_{\nu} = 
    \nu' \pi \left(\frac{R_\text{b}}{D_\text{L}} \right)^2 I_{\nu'},
\end{equation}
where $\nu' = (1+z)\nu$, $z=0.306$ is the redshift of the blazar OJ 287, $D_\text{L}=1602$ Mpc its luminosity distance, and $I_{\nu'}$ 
is the specific intensity (calculated for $\nu'$) of the thermal bremsstrahlung radiation
at the outer boundary of the outflow, i.e. at $R = R_\text{b}$.
The spectrum due to optically thin bremsstrahlung emission can be obtained using
\begin{equation}
\label{Inu_thin}
    I_\nu = (4/3)R_\text{b} j_\nu,
\end{equation}
where $j_\nu$ is the thermal bremsstrahlung emission coefficient which here is taken as
\begin{equation}
    \label{j_nu}
    4 \pi j_\nu  = 6.8 \times 10^{-38} n_\text{b}^{2} T_\text{b}^{-1/2} \exp\left\{-\frac{h\nu}{k T_\text{b}}\right\}
    \mbox{erg}\,\mbox{s}^{-1}\,\mbox{cm}^{-3}\,\mbox{Hz}^{-1}.
\end{equation}
The spectrum given by equation (\ref{Inu_thin}) is correct as long as the medium 
producing the radiation is optically thin.
For the outflow bubble defined by condition (\ref{cond_thin}),
this turns out to be the case at optical frequencies.
However, this is not the case for lower frequencies where the emission is attenuated
by self-absorption. To account for this effect present at low frequencies, 
one can alternatively employ the specific intensity:
\begin{align}
\label{Inu_self}
    I_\nu &= B_\nu \left(
    1 - \exp\{-\tau_\nu \}
    \right),\\
    \label{tau_nu}
    \tau_\nu & = 
    \int_0^{4R_\text{b}/3}
    j_\nu/B_\nu ds
    \approx
    (4/3)R_\text{b} j_\nu/B_\nu,
\end{align}
which is thermal bremsstrahlung emission with self-absorption
given by Kirchhoff's law and the black-body spectral
radiance of temperature $T_\text{b}$:
\begin{equation}
\label{B_nu}
    B_\nu = \frac{2 h  \nu^3}{c^2}
    \left(
    \exp \left\{\frac{h\nu}{k T_\text{b}}\right\} - 1
    \right)^{-1}. 
\end{equation}

We consider the photon field density $n_\text{ph}$ generated by thermal bremsstrahlung, as the
target photon field for inverse Compton scattering of secondary electrons within the outflow
(this radiation process is discussed in Section~\ref{subsec:ppemit}).
We calculate
this target photon field density (number of photons per unit
energy, per unit volume) as
\begin{equation}
\label{n_ph}
    n_\text{ph}(\epsilon, T_\text{b}) = \frac{4\pi I_\nu(R_\text{b},T_\text{b})}{c h \epsilon},
\end{equation}
where $\epsilon = h\nu$ is the energy of the thermal bremsstrahlung photons.
We also employ the photon field density of equation (\ref{n_ph}) to calculate
the attenuation of the $\gamma$-rays due to photon-photon annihilation,
as described in the following subsection.

\begin{figure}
	\includegraphics[width=8.5cm]{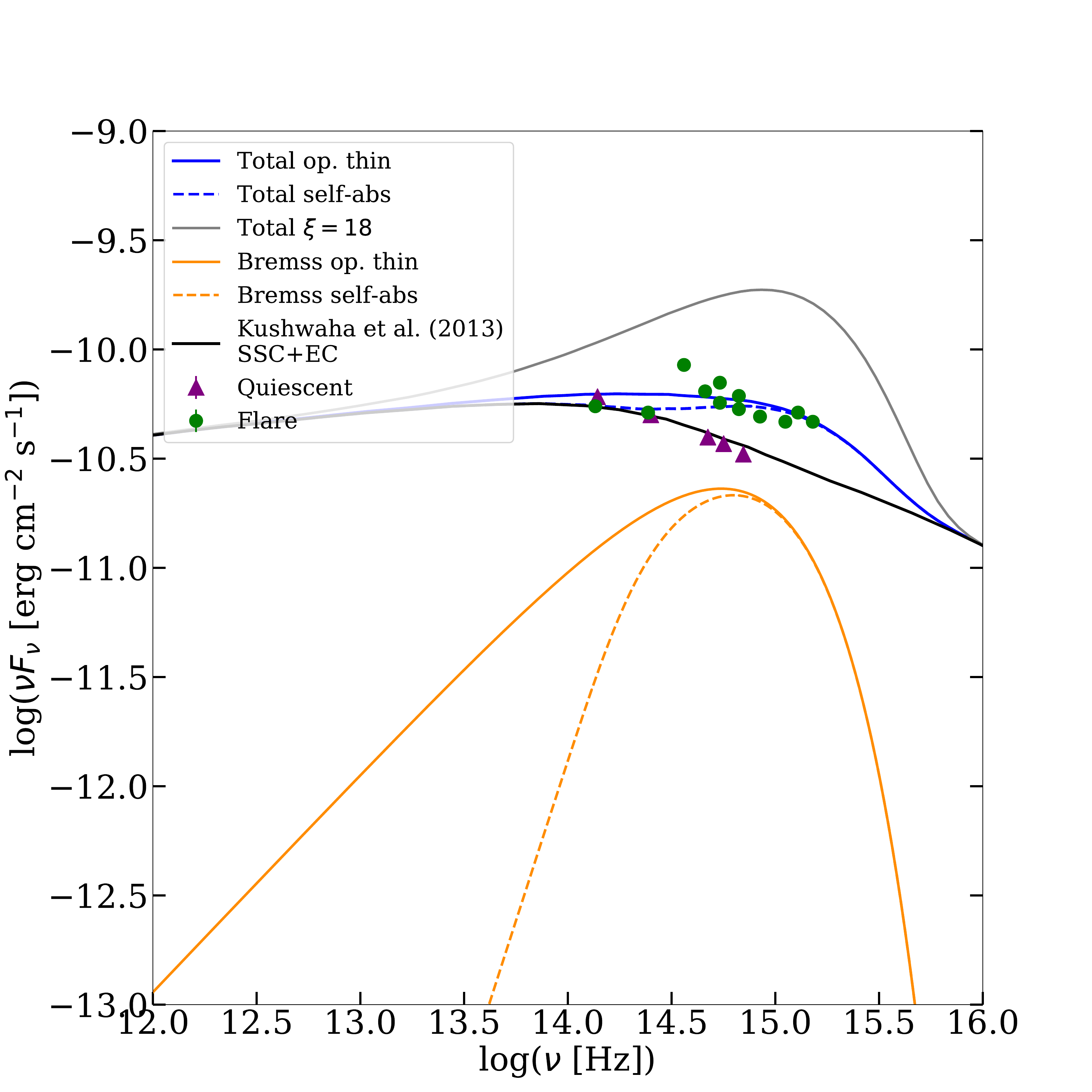}
    \caption{
    SED of the 2015 major flare (green data points) of OJ 287 at NIR-UV bands
    (adapted from \citealt{2018MNRAS.473.1145K}).
    The magenta data points correspond to a previous quiescent state
    and the black curve is a jet synchrotron emission model
    (both data and curve adapted from \citealt{2013MNRAS.433.2380K}).
    See the text for the description of the different flare emission curves.}
\label{fig:V19}
\end{figure}

We note that in all the models considered in this paper, there is no substantial difference
between self-absorbed (equation~\ref{Inu_self}) and optically thin 
bremsstrahlung emission (equation~\ref{Inu_thin}) 
at optical energy bands.
For the parameters considered here, the spectra of these two approaches
start to diverge for energies $\lesssim kT_\text{b}$.
An example that illustrates this is shown in Figure~\ref{fig:V19},
where the orange, solid curve is calculated using equation (\ref{Inu_thin}) (optically thin bremsstrahlung)
and the orange, dashed curve is obtained using equation (\ref{Inu_self}) (bremsstrahlung with self-absorption).
Whereas this discrepancy is not substantial to model the optical outburst data,
the difference between the two approaches can be noticeable in the resulting X-ray spectrum
of relativistic electrons Comptonising
the bubble radiation field
(this HE process is described in Section~\ref{subsec:ppemit}).
Since an optically thin bremsstrahlung spectrum is not physically possible
for photons at arbitrary low energies, in this work we adopt the thermal spectrum
given by equation~({\ref{Inu_self}}).
The spectrum that results from this bremsstrahlung self-absorbed emission is compatible
with the thermal spectrum of BH-disc impact outflows proposed by \cite{2016MNRAS.457.1145P}.

The time that the spherical outflow takes to expand from the initial radius
$R_0$ to  the outburst radius $R_\text{b}$ 
depends on the dynamics of the expansion.
Here we follow the expansion dynamics proposed by \cite{1996ApJ...460..207L}.
In this approach, the outer boundary of the outflow expands with a
velocity equivalent to the speed of sound of the bubble internal gas,
which is radiation pressure dominated.
With this assumption
the velocity of the outflow's outer boundary is
\begin{equation}
\label{VsLV}
    c_\text{s} = c_\text{s0}\left(R /R_0\right)^{-1/2},
\end{equation}
the radius of the outflow evolves with time as 
\begin{equation}
    R(t) = R_0 \left(1+\frac{3}{2}\frac{c_\text{s0}}{R_0}t\right)^{2/3},
\end{equation}
and the time that the bubble takes to attain the outburst radius $R_\text{b}$ is
\begin{equation}
\label{DtbLV}
    \Delta t_\text{b} = \frac{2}{3}\frac{R_0}{c_\text{s0}}\left(
    \xi^{3/2} -1
    \right).
\end{equation}
In equations (\ref{VsLV})-(\ref{DtbLV}), $c_\text{s0} = \sqrt{8/7} v_\text{r}$ is the initial speed of sound 
of the gas within bubble and $v_\text{r}$ is the velocity of the accretion disc material in the co-moving 
frame of the secondary BH at the impact event
\citep{1996ApJ...460..207L,2016MNRAS.457.1145P}.

In short, in this work 
we consider a BH-disc impact scenario in which 
(i) the impact produces two outflows, one above and the other below the
accretion disc
(as proposed by \citealt{1998ApJ...507..131I} and \citealt{2016MNRAS.457.1145P}),
(ii) we attribute the observed emission to the outflow 
emerging in the direction of the observer only, and
(iii) this outflow expands as described in 
\cite{1996ApJ...460..207L}, \cite{2018ApJ...866...11D}, and \cite{2019ApJ...882...88V}.
Within this approach, we employ the condition
(\ref{cond_thin}) to define the state of the 
expanding spherical outflow 
(e.g., radius, gas density, temperature)
from which we calculate  optical, X-ray, and $\gamma$-ray
emission to model the MW 2015 flare data
(see Section~\ref{sec:SED_models}).

This is an idealised and simplified 
emission scenario.
As discussed by \cite{2016MNRAS.457.1145P}, outflows
driven by BH-disc impacts might be far from
having uniform internal structure.
In addition, the vertical stratification of the disc 
and the gravitational influence of the primary SMBH
would lead to a non-spherical outflow morphology.
Regarding the opacity of the impact outflow, Kramer's formula 
(\ref{kappa_a}) for the opacity of free-free and bound-free transitions
is a very crude approximation
for the absorption opacity when compared to more accurate numerical
computations for static stellar interiors
\citep{1996ApJ...464..943I,2005ApJ...623..585F}.
Furthermore, the impact outflow discussed here is
not static and the expanding nature of the emitting plasma
introduces effects
\citep{2014ApJ...787L...4S, 2016MNRAS.457.1145P},
not quantified in the present emission model.

Surprisingly, despite neglecting the aforementioned
physical effects, the emitting volume proposed by \citep{1996ApJ...460..207L}
appears to explain several aspects of the observed recurrent optical flares,
not successfully explained by alternative models (see the introduction and
references therein).
Particularly, to explain the origin of the 2015 major 
optical outburst, \cite{2019ApJ...882...88V} 
consider the BH masses $M_1= 1.835\times10^{10}$M$_\odot$
and $M_2=1.5\times10^{8}$M$_\odot$
for the primary and secondary BHs, respectively,
an impact distance of $R_{\text{imp}}\sim17500$ AU 
from the primary BH where
$n_\text{d} \sim 1.75\times 10^{14}$ cm$^{-3}$, $v_\text{r} \sim 0.12c$,
and an outflow bubble with initial radius $R_0 \sim 47$ AU.
They follow \cite{1996ApJ...460..207L} to define the transparency condition
of the emitting bubble (see equation \ref{cond_thin}) assuming
 $X=0.85$, $Z=0.02$ for the mass fractions of hydrogen and metals. 
If one considers the above parameters and the guillotine factor $t_\text{bf} = 25$
for the absorption opacity (see equation~\ref{K_a}), 
the flux that results due to optically thin, thermal bremsstrahlung radiation 
employing equations (\ref{nuFnu})-(\ref{j_nu})
matches well with the optical data of the 2015 flare as shown in the
blue solid curve in Figure~\ref{fig:V19}.
This thermal flux corresponds to a bubble 
that expands a factor of $\xi\sim 31$
(and not $\xi\sim 18$ which is perhaps a typo in the text of
\citealt{2019ApJ...882...88V}),
and has a temperature
$T_\text{b}\sim3.4\times10^4$ K and gas number density 
$n_\text{b} \sim 4.1\times10^{10}$ cm$^{-3}$
when it let the radiation escape freely
(using $\xi =18$ the calculated flux, which is plotted by the grey 
curve in Figure~\ref{fig:V19}, overshoots the data considerably).

In this paper we investigate whether a BH-disc impact is a viable scenario
to explain the simultaneous spectral changes in the broadband SED of OJ 287.
We then look for the conditions that allow the outflow bubble described in this Section
to account simultaneously for the optical, X-ray and $\gamma$-ray excess 
(we describe the HE emission model in Section~\ref{sec:non-thermal}).
To do this, we  use $v_\text{r}$, $f_\text{R}$ and $n_\text{d}$
as free parameters to obtain outflow bubbles with different properties
(such as $R_\text{b}$, $n_\text{b}$, and $T_\text{b}$)
which thermal emission is consistent with the optical data of the 2015 outburst.
The masses of the SMBHs,
primary $M_1 =1.835\times10^{10}$M$_\odot $ and secondary $M_2 =1.5\times10^{8}$M$_\odot$ 
\citep[taken from ][]{2018ApJ...866...11D},
are considered fixed in this work.
We also use the fixed values of $X=0.85$ and $Z=0.02$
for the mass fraction of hydrogen and metals
 \citep[following][]{2019ApJ...882...88V}, and
$t_\text{bf} = 5$ for the guillotine factor of absorption opacity within the outflow bubble.
These values appear to be appropriate for the 
bubble models considered here and also
provide a good agreement with the data.

\subsection{The opacity of the impact outflow to gamma-ray photons}
\label{subsection:photon-photon}
The flux of potential  $\gamma$-rays produced in the impact outflow is
susceptible to be attenuated by internal absorption.
We consider the thermal bremsstrahlung radiation
discussed in this Section as the dominant source of soft photons for 
$\gamma$-ray annihilation.
If $L^0(E_{\gamma})$ is the luminosity of $\gamma$-ray photons 
of energy $E_\gamma$ produced in the 
impact outflow, the luminosity of $\gamma$-rays that escape the emission
region can be calculated as $L(E_\gamma) = L^0(E_\gamma) 
\exp\{-\tau_{\gamma\gamma}(E_\gamma)\}$, where $\tau_{\gamma\gamma}$
is the optical depth of photon-photon annihilation.

To calculate $\tau_{\gamma\gamma}$ we assume for simplicity that
the thermal bremsstrahlung photon field is isotropic and uniform within the outflow volume and non-effective for photon-photon collisions outside this volume. 
Thus, the opacity due to $\gamma$-ray absorption can be calculated as
\begin{equation}
\label{tau_gg}
    \tau_{\gamma\gamma}(E_\gamma) = \int_0^{l_{\gamma\gamma}}ds
    \int_{m_\text{e}^2 c^4/E_\gamma}^{\infty} d\epsilon\,
     \sigma_{\gamma\gamma}(E_\gamma, \epsilon) n_\text{ph}(\epsilon, T_\text{b}) \,,
\end{equation}
where, 
$n_\text{ph}$ is the photon field of thermal bremsstrahlung radiation given by equation (\ref{n_ph}),
$l_{\gamma\gamma}$ is the length of the path that $\gamma$-rays photons travel before
leaving the outflow volume,
and 
\begin{equation}
    \sigma_{\gamma\gamma}(E_\gamma,\epsilon) = 
    \frac{\pi r_\text{e}^2}{2}(1-\beta^2)
    \left[
    (3-\beta^4)\ln\left(\frac{1+\beta}{1-\beta} \right)
    + 2\beta(\beta^2-2)
    \right],
\label{sigma_gg}
\end{equation}
is the total cross section for
photon-photon collisions (e. g., \citealt{1985Ap&SS.115..201A},
\citealt{2010A&A...519A.109R}), where $\beta^2 =1- m_\text{e}^2c^4 /(E_\gamma\epsilon)$, and $r_\text{e}$ is the classical electron radius. 
The main uncertainty of this approach is the size of the length 
$l_{\gamma\gamma}$, which 
depends on the direction of the line of sight as well as on the morphology of
the emission region (see Figure~\ref{fig:out-shock}b).

\begin{figure}
	\includegraphics[width=9.5cm]{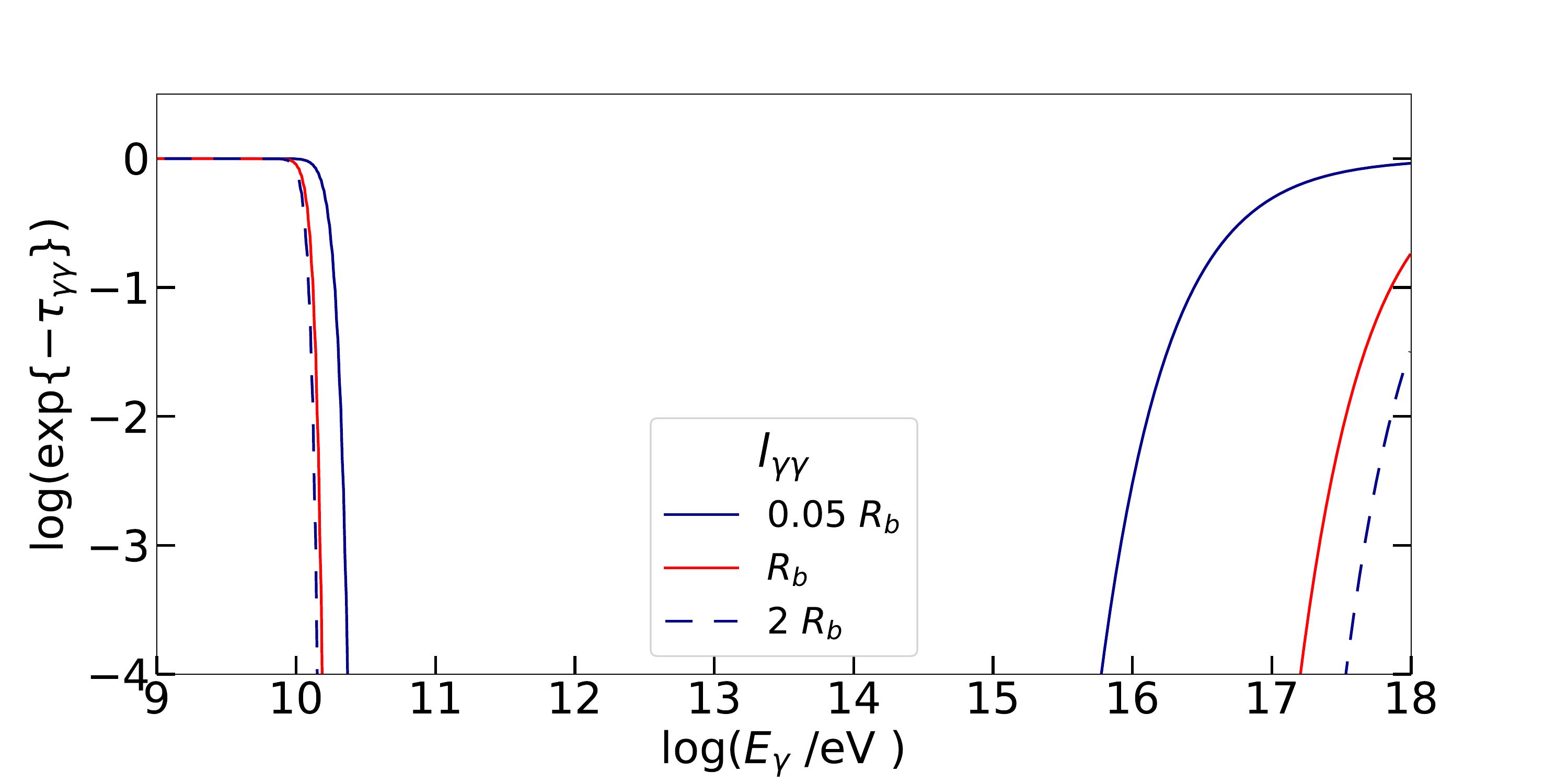}
    \caption{
    Attenuation factor due to internal $\gamma$-ray annihilation 
    in the impact outflow  (see the text).
    The different curves correspond to different lengths assumed for the
    $\gamma$-ray path before leaving the absorption region.
    }
    \label{fig:tau_gg}
\end{figure}

We show in Figure~\ref{fig:tau_gg} the attenuation factor $\exp\{ -\tau_{\gamma\gamma} \}$ 
calculated for 
$l_{\gamma\gamma}= 0.05R_\text{b}$ and $2R_\text{b}$,
(blue solid and dashed curves respectively).
Clearly, the difference between these two extreme cases
is not substantial for defining the cut-off energy at $\sim 30$ GeV
of the resulting $\gamma$-ray flux.
The attenuation factors calculated with $l_{\gamma\gamma}= 0.05R_\text{b}$
and $2R_\text{b}$ are drastically different for $\gamma$-rays above $10^{15}$ eV.
However, this difference is irrelevant for all the emission models 
derived in this work as the maximum energy of the accelerated CR protons 
(which potentially produce HE $\gamma$-rays)
is bellow $\sim 10^{12}$ eV (see subsection \ref{subsec:ppemit}).
The attenuation factors displayed in Figure~\ref{fig:tau_gg} 
are calculated with the parameters of the model M2, specified in Table 1 and 
we note the same behaviour discussed above in all the emission models.
Thus, for simplicity, we adopt the intermediate value of
$l_{\gamma\gamma} = R_\text{b}$ for the $\gamma$-ray attenuation in all the
SED models derived in Section~\ref{sec:SED_models}.

 $\gamma$-rays fluxes can also be attenuated by the extra-galactic background light (EBL)
on the way to the Earth.
For a source of red-shift like that of the blazar OJ 287, 
photon-photon annihilation by the EBL is significant for $\gamma$-rays with 
energies
$\gtrsim 100$ GeV \citep{2010ApJ...712..238F}. 
Since the internal absorption in the emission model discussed here
let escape photons with energies $\lesssim 50$ GeV (see Figure~\ref{fig:tau_gg})
we neglect attenuation by the EBL in the SED models derived in Section~\ref{sec:non-thermal}.

\section{The outflow non-thermal emission}
\label{sec:non-thermal}

\subsection{Conditions for shock formation}

\label{subsec:shock}

Depending on the location of the secondary SMBH impact, the resulting outflow may expand inside or 
outside a corona of hot gas that surrounds the primary SMBH (see Figure \ref{fig:out-shock}a).
If the impact takes place within the coronal region, the resulting outflow 
is unlikely to drive a strong shock as the speed of sound of the coronal gas may be comparable to
the expansion velocity  of the impact 
outflow\footnote{
The electron temperature of an AGN corona may be $T_\text{c,e}\sim 10^{8}$ K \citep{2015MNRAS.451.4375F}.
The proton temperature $T_\text{c,p}$ in the corona may be two or three orders 
of magnitudes higher.
Considering the temperature of protons, the speed of sound of an isothermal corona
may then be of $\sim 0.095c \,(T_\text{c,p} / 10^{11}\mbox{K})^{1/2} $ which is comparable to the expansion velocity $c_\text{s}$ of the impact outflow (see equation \ref{VsLV}).
}. 
The scale height of this putative hot structure surrounding the central BH of AGNs, though model dependent,  
may be of $\sim$ 10-20 $R_\text{g}$ \citep{2015ApJ...802..113K,2015MNRAS.451.4375F,2017ApJ...847...96L},
with $R_\text{g}=GM/c^2$.
For the OJ 287 November 2015 outburst, the SMBH binary model predicts the 
impact of the secondary SMBH at a distance of $R_\text{imp} = 17566$ AU $\sim100 R_\text{g}$ 
from the primary SMBH \citep{2018ApJ...866...11D},
which is outside the central hot corona.
Other impact events like those corresponding to the outburst epochs of 1994 and 2005 
are predicted closer to the central corona.

 Outside the coronal region, the impact outflow may interact with a high velocity AGN wind (driven by the primary SMBH). 
 Observational evidences 
 \citep[e.g.][]{2012MNRAS.426..656S,2013MNRAS.429.1872C}
 as well as theoretical studies
 \citep[][and references therein]{2015ApJ...812...90M, 2019A&A...630A..94G}
indicate that AGN winds take place in the vicinity of the central engine of AGNs
with velocities representing a significant fraction of the speed of light.

A strong shock that accelerate particles can then be formed due to the interaction
of the AGN wind and the expanding impact outflow, provided that
\begin{equation}
\label{M_sA}
    \mathcal{M}_\text{sA} = V_\text{s0}
    \left(c_\text{w}^2 + c_\text{A}^2\right)^{-1/2}>>1,
\end{equation}
where the $\mathcal{M}_\text{sA}$ is the magnetosonic Mach number,
$V_\text{s0} =c_\text{s} + \cos(\theta)V_\text{w}$ is the velocity of the AGN wind material in the rest frame of 
bubble expanding front, 
$V_\text{w}$ and $c_\text{s}$ are the AGN wind and the bubble expansion velocities, respectively (see equation~\ref{VsLV}),
in the frame of the primary SMBH,
$\theta$ is the angle between the wind velocity and the vector $-\hat{n}$,
being $\hat{n}$ the unit vector 
normal to the shock surface (see Figure~\ref{fig:out-shock}b),
and $c_\text{w}$ and $c_\text{A}$ are the sonic and Alfv\'enic speeds, of the AGN wind respectively.
Using $c_\text{w}^2 = \gamma_\text{w} k T_\text{w} / m_\text{p}$ and 
$c_\text{A}^2 = B_\text{w}^2 / (4\pi m_\text{p} n_\text{w})$ where
$T_\text{w}$ 
$n_\text{w}$, and
$B_\text{w}$
are the temperature, number density, and magnetic field of the AGN wind,
the magnetosonic Mach number can be estimated as:
\begin{align}
\label{M_SA}
\nonumber
    \mathcal{M}_\text{sA} = & V_\text{s0} \left( \frac{m_\text{p}}{\gamma_\text{w} k T_\text{w}} \right)^{1/2}
    \left( 1 + \frac{B_\text{w}^2}{4\pi\gamma_\text{w} k T_\text{w} n_\text{w}} \right)^{-1/2} \\
    \nonumber
     \sim & 13.9 \left(\frac{V_\text{s0}}{0.2 c}\right)\left(\frac{10^9 \text{K}}{T_\text{w}}\right)^{1/2}\, \times \\
     & \left[
     1 + 0.3 
     \left( 
     \frac{B_\text{w}}{1\mbox{G}}\right)^2 
     \left( \frac{10^9 \mbox{K}}{T_\text{w}} \right)
     \left( \frac{10^6\,\mbox{cm}^{-3}}{n_\text{w}} \right)
     \right]^{-1/2},
\end{align}
where we use $\gamma_\text{w}=5/3$.

If the shock is radiation-dominated, the appropriate specific heat ratio for the gas
within the shock is $\gamma_{\text{sh}}=4/3$ and the maximum shock compression is 7.
Alternatively, if the shock is optically thin, radiation fields 
(originated within the shell or externally)
produce no substantial effects on the properties of the shocked gas.
In this case 
$\gamma_{\text{sh}}=5/3$ is the appropriate value for the specific heat ratio
which leads to a maximum compression of 4.
The shocked material is optically thin as long as the time scale of photon diffusion
$t_{\text{diff}} = \tau R_\text{b} / c =  \sigma_\text{T} n_{\text{sh}} R_\text{b}^2 /c$
(i.e., the time that photons take to leave the the swept up shell) is much shorter than the
time scale of the outflow expansion $t_{\text{exp}} \sim R_\text{b} / v_\text{r}$ 
(here we consider the initial  outflow velocity $\sim v_\text{r}$ as the maximum velocity of the expansion, see equation \ref{VsLV}).
Thus, the shock formed by  the interaction of the AGN wind and the outflow bubble
is optically thin if gas number density of the shocked gas is
\begin{equation}
    n_{\text{sh}}<< \frac{c}{v_\text{r} \sigma_\text{T} R_\text{b}} \sim 5 \times 10^8 \text{cm}^{-3} 
    \left(
    \frac{0.1 c}{v_\text{r}}
    \right)
    \left(
    \frac{0.01\text{pc}}{R_\text{b}}
    \right).    
    \label{n_sh}
\end{equation}
As we will see in Section~\ref{sec:SED_models}, in all the emission models considered in this paper
the parameters of wind-outflow shock fall in the optically thin regime.

We compare the associated mechanical luminosity of the AGN wind with the mass accretion
power of the primary SMBH by defining the wind efficiency parameter
\begin{equation}
\label{eta_w}
    \eta_\text{w} \equiv  \frac{\pi R_\text{imp}^2 \rho_\text{w} V_\text{w}^3}{\dot{M}_1 c^2}
    = 0.28 \left(\frac{n_\text{w}}{5\times10^6\,\mbox{cm}^{-3}}\right)
     \left(\frac{V_\text{w}}{0.25 c}\right)^3.
\end{equation}
In this ratio, we use $R_\text{imp}=17566$ AU for the distance of the secondary SMBH-disc impact
(corresponding to the 2015 outburst, see \citealt{2018ApJ...866...11D})
and $\dot{M}_1 = 0.12 \dot{M}_\text{Edd}$ for the mass accretion rate of the primary SMBH 
of OJ 287 \citep{2019ApJ...882...88V}.

According to the estimations discussed in this subsection, 
a strong shock can be formed due to the interaction of the AGN wind and the impact outflow,
if the AGN wind has velocity, temperature, magnetic field, and gas density constrained to values
$V_\text{w}\gtrsim 0.2 c$,
$T_\text{w}\lesssim 10^{9} $ K,
$B_\text{w}\lesssim 1$G, 
and $n_\text{w}\gtrsim 10^{6} $ cm$^{-3}$, respectively.
In the  next subsection, we  assume that such an AGN wind exists above the accretion disc at the 
location of the secondary SMBH impact and then we derive the associated non-thermal, hadronic
emission of the  accelerated CR protons.

We consider the total energy of the emitting CR protons to be a small fraction of the
the kinetic wind energy $E_\text{w}$ that crosses
the shock formed due to the interaction of the AGN wind and the 
expanding outflow (see equation \ref{normJp} in the next subsection).
We calculate the impinging wind energy $E_\text{w}$ assuming for simplicity 
an AGN wind with local plane-parallel geometry of uniform density $\rho_\text{w}$
and velocity $V_\text{w}$.
Considering the surface of the forward shock as nearly spherical,
the flux of wind kinetic energy impinging
on the shock surface can be estimated as
\begin{equation}
    \frac{dE_\text{w}}{dtdA} = \frac{1}{2}\rho_\text{w} \left[c_\text{s} + \cos(\theta)V_\text{w}\right]^3,
\end{equation}
where $\theta$ is the angle between the wind velocity and the vector
$-\hat{n}$, being $\hat{n}$ the unit vector normal to the surface of the shock
(see Figure~\ref{fig:out-shock}b).
Thus, the wind kinetic energy that crosses the shock front
during the time $\Delta t_\text{b}$
(which corresponds to the period when the outflow bubble expands from the
$R_0$ to $R_\text{b}$), can be calculated as
\begin{equation}
\label{E_w}
    E_\text{w}  = \int_{0}^{\Delta t_\text{b}} dt R(t)^2 
    \int_0^{2\pi}d\phi 
    \int_0^{\pi/2} d\theta \sin(\theta) 
    \frac{d E_\text{w}}{dt dA}.
\end{equation}
Integrating equation (\ref{E_w}) with $\Delta t_\text{b}$, $c_\text{s}(t)$, 
and $R(t)$ given by equations (\ref{VsLV})-(\ref{DtbLV})
(corresponding to an expanding bubble according to \citealt{1996ApJ...460..207L})
gives
\begin{align}
\nonumber    E_\text{w} = \pi \rho_\text{w} R_0^3 & \left[ 
    \frac{c_\text{s,0}^2}{2}\left(\xi^{2} -1\right)+
    \frac{3}{5}c_\text{s,0}V_\text{w}\left(\xi^{5/2} -1\right)+ \right. \\
    & \;\;\;\;\frac{V_\text{w}^2}{3}\left(\xi^{3} -1\right)+
    \left.\frac{V_\text{w}^3}{14 c_\text{s,0}}\left(\xi^{7/2} -1\right) \right].
\end{align}
$R_0$ and $\xi$ and $c_\text{s,0}$ can be obtained as described in Section~\ref{sec:scenario}.
We employ $\rho_\text{w}$ and $V_\text{w}$ as free parameters that are found
by matching the calculated emission of hadronic origin to the observed X-ray and 
$\gamma$-ray data (see the Section~\ref{sec:SED_models}).
In reality, the shock formed due to the collision of the outflow bubble and the AGN wind
may follow a bow-shock morphology (as depicted in Figure~\ref{fig:out-shock}b).
Therefore, equation (\ref{E_w}) slightly underestimates the wind kinetic energy
impinging on the shock surface.

\subsection{Emission from proton-proton interactions }
\label{subsec:ppemit}

To calculate the potential non-thermal radiation produced by the impact outflow, we consider acceleration
of CRs  in the shock
formed by the interaction of the expanding bubble with the AGN wind driven by the primary SMBH
(see Figure \ref{fig:out-shock}). Assuming diffusive shock acceleration (DSA),  the
acceleration rate of CR protons can be written as
\begin{equation}
    t_\text{acc}^{-1} = \frac{1}{E}\frac{dE}{dt} = \frac{V_\text{s0}^2}{D(E)},
\end{equation}
where $V_\text{s0}$ is the upstream velocity in the co-moving frame of the shock
(see the previous subsection), and $D$ is the CR diffusion coefficient in the acceleration region.
Since super-Alfv\'enic turbulence is likely to develop in 
the super-Alfv\'enic, supersonic AGN wind,
we adopt for simplicity a spatially uniform,
Kolmogorov-like diffusion coefficient of the form
\citep{2006ApJ...642..902P,2019MNRAS.490.4317C}:
\begin{equation}
\label{D}
    D = D_0 \left(\frac{E}{E_0}\right)^{1/3}
    \left(\frac{B}{B_0}\right)^{-1/3}
\end{equation}
with $E_{0} = m_\text{p}c^2 + 2 m_\pi c^2 + m^2_\pi c^4 /(2m_\text{p}c^2) = 1.22$ GeV (the threshold energy for the production of $\pi^0$ mesons), $B_0=1$ G, and
we choose the normalisation constant  $D_0 = 5\times 10^{25}$ cm$^2$ s$^{-1}$.
This form of the diffusion coefficient is motivated by the condition
\begin{equation}
\label{cond_diff}
    \sqrt{2D \Delta t_\text{b}}>\Delta R,
\end{equation}
in which CRs protons with energies $>E_0$ diffuse
from the forward shock into the bubble   
in a time $< \Delta t_\text{b}$.
Here $\Delta t_\text{b}$ is the time that the bubble
takes to produce the optical outburst 
(see equation~\ref{DtbLV}), 
and 
$\Delta R\sim 0.5 R_\text{b}$
is the thickness of the 
shell\footnote{
For a plane-parallel wind impinging on a spherical surface
with sonic Mach number $>>1$,
the thickness of the swept up shell at $\theta = 0$ (see Figure~\ref{fig:out-shock}b)
can be well approximated as
$\Delta R \sim  \epsilon (0.76 + 1.05\epsilon^2)R_\text{b} \sim 0.2 R_\text{b}$, 
where  $\epsilon= (\gamma -1 )/ (\gamma + 1 )=1/4$, for $\gamma = 5/3$ 
\citep[][their equation  22]{2003JGRA..108.1323V}.
At $\theta = \pi/2$, this thickness is $\Delta R \sim 0.75 R_\text{b}$ \citep[][their Figure~4]{2003JGRA..108.1323V}.
In the condition (\ref{cond_diff}), we employ the intermediate value of 
$\Delta R = 0.5 R_\text{b}$
}
of shocked AGN wind material (region ii in Figure~\ref{fig:out-shock}b).
With this normalisation, for $B=1$~$\mu$G and $E = 10$ GeV, equation (\ref{D})
gives $D\sim10^{28}$ cm$^{2}$ s$^{-1}$ which, coincidentally, is of the order of the average diffusion coefficient inferred for our Galaxy.

Given the radius $R_\text{b}$, density $n_\text{b}$, temperature $T_\text{b}$,
and photon field $n_\text{ph}$ of the outflow bubble,
the magnetic field $B$ in the acceleration region required to accelerate CR protons
up to  a maximum energy $E_\text{max}$ can be found by balancing the acceleration
rate $t^{-1}_\text{acc}$ with the rates of energy losses of protons in the swept up shell 
(region ii in Figure \ref{fig:out-shock}b):
\begin{equation}
\label{rate_balance}
    t^{-1}_\text{acc} (E_\text{max},B) =
    t^{-1}_\text{diff}(E_\text{max},B) + t^{-1}_\text{pp}(E_\text{max},n_\text{b}) 
    + t^{-1}_{\text{p}\gamma}(E_\text{max},n_\text{ph}). 
\end{equation}
In this equation, we consider the energy loss rates due to   
CR diffusion $t^{-1}_\text{diff}$, 
proton-proton interactions $t^{-1}_\text{pp}$  
(of CRs with the thermal protons in the swept up shell), 
and photo-pion production $t^{-1}_{\text{p}\gamma}$ 
(due to interactions of CRs with the thermal bremsstrahlung radiation of the expanding bubble, see equation \ref{n_ph}).
We note that these cooling rates change as the outflow bubble evolves. 
For the sake of simplicity, we calculate these cooling rates at the time when the outflow bubble allows the
thermal bremsstrahlung photons to escape (see Section \ref{sec:scenario}).
The expressions that we employ for the rate terms in equation (\ref{rate_balance}) are described
in Appendix  (\ref{appendix:rates_protons}).
In Figure \ref{fig:rates}a, we plot the characteristic times
as a function of the proton energy
of the acceleration and energy loss rates of protons in 
the swept up shell with parameters corresponding to 
a particular emission model derived in Section~\ref{sec:SED_models} 
(the model M2, see also Table~1).

Considering that CRs protons escape isotropically from their acceleration zone,
we note that the material of the outflow bubble (see Figure \ref{fig:out-shock}b) 
is the main target for $p$-$p$ interactions (of CR protons with the local thermal ions).
For the parameters of the emission models considered here, 
these interactions occur much more efficiently
within the outflow bubble than within the shell of swept up material.
This can be seen by comparing the $p$-$p$ cooling time curves (orange)
in the upper and lower panels of Figure~\ref{fig:rates}. These curves
are the $p$-$p$ cooling times in the shell of shocked AGN wind
and in the region within outflow bubble, respectively.
For an AGN wind with gas density $n_\text{w} \sim 5 \times 10^{6}$ cm$^{-3}$, for instance,
CR protons with energies of $\sim$1 TeV cools via
$p$-$p$ interactions within the swept up shell
in a time scale of years.
By contrast, the $p$-$p$ cooling time of CR protons within an outflow bubble
of gas density $\sim 10^{10}$ cm$^{-3}$ is of few days.

The neutral and charged pions ($\pi^0$ and $\pi^{\pm}$) produced out of $p$-$p$ interactions decay into $\gamma$-rays and  electron-positron pairs $e^{\pm}$ through the channels:
\begin{align}
\label{pi0gamma}
&\pi^0 \rightarrow \gamma + \gamma,\\
&\pi^{\pm} \rightarrow \mu^{\pm} + \nu_\mu(\bar{\nu}_\mu),\\
  \label{pipmpairs}
 &\mu^{\pm}\rightarrow e^{\pm} +\bar{\nu}_\mu(\nu_\mu) + \nu_e(\bar{\nu}_e),
\end{align}
where $\mu^{\pm}$ and $\nu_s$ represent muons and neutrinos, respectively. 
For the time scales of the problem discussed here, we can assume that pions and muons decay
instantaneously in the primary SMBH rest frame.
In all the emission models derived in Section~\ref{sec:non-thermal}, 
the secondary $e^{\pm}$ pairs cool down more efficiently due
to IC scattering (of the thermal bremsstrahlung radiation field generated by the outflow bubble)
than due to synchrotron radiation. This is illustrated with the cooling times of $e\pm$ 
displayed in Figure~\ref{fig:rates}b, corresponding to the emission model M2 
(specified in Table 1).

\begin{figure}
\centering
	\includegraphics[width=8.5cm]{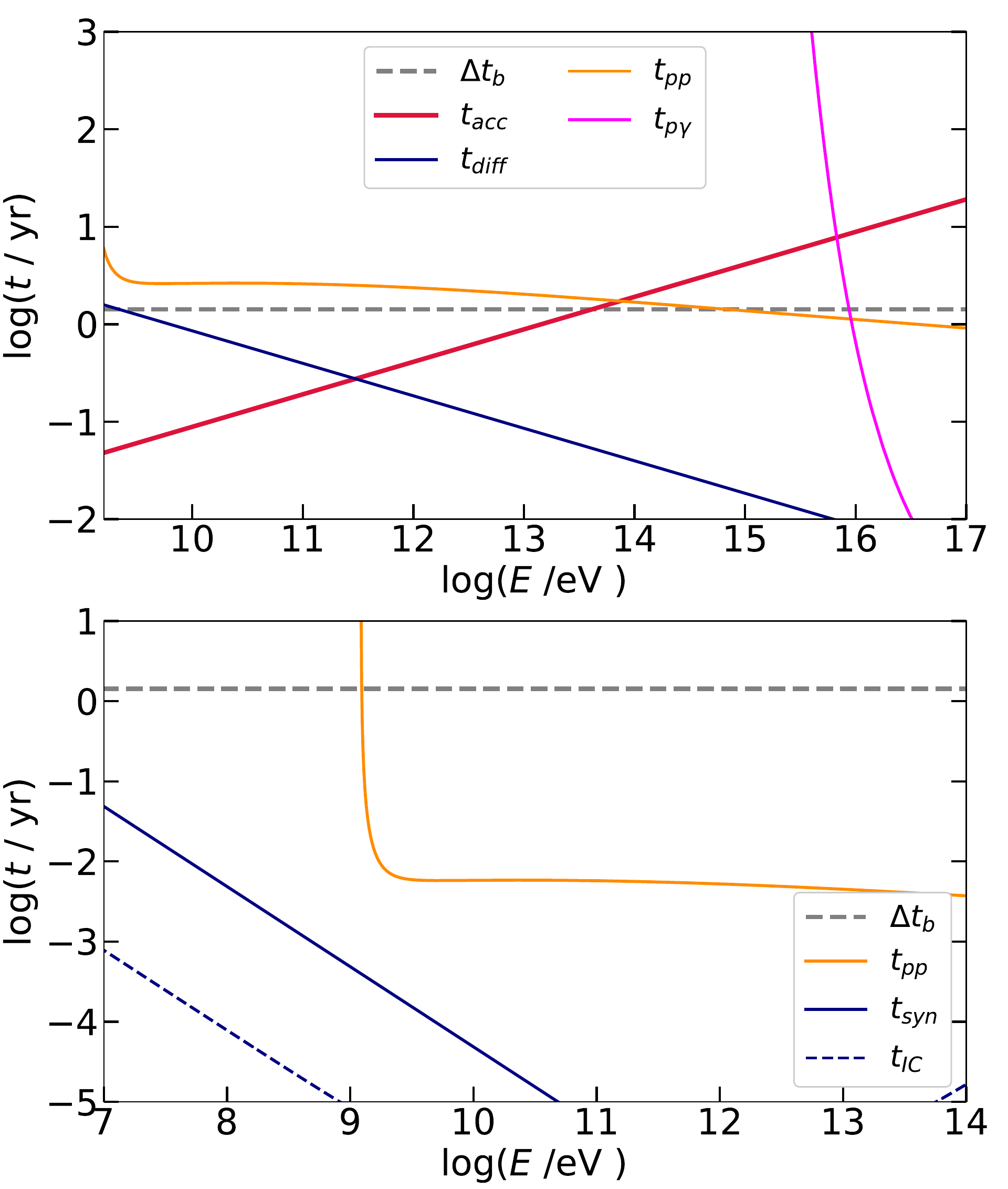}
    \caption{
    Upper: Characteristic times of acceleration and cooling for CR protons in the shell of swept up material
    (region ii in Figure~\ref{fig:out-shock}b) driven by the impact outflow (see the text).
    Lower: Cooling times for protons and secondary $e^{\pm}$ pairs
    within the outflow bubble (region iii in Figure~\ref{fig:out-shock}b; see the
    text). The curves in the these panels corresponds to the parameters of the emission model M2 (see Table 1).
 }
    \label{fig:rates}
\end{figure}

To calculate the emission due to $\pi^{0}$ decay as well as due to the secondary $e^{\pm}$ pairs generated out of $p$-$p$ interactions, we first assume that a stationary population of CRs has been 
injected within the volume of the outflow bubble.
The volume that this CR population occupies within the bubble (region iii in
Figure~\ref{fig:out-shock}b) can be estimated
according to the distance that CRs protons diffuse in a time $\Delta t_\text{b}$
(the time that the outflow bubble takes to manifest as a flare).
Employing the diffusion coefficient defined in equation~(\ref{D}), 
the distance that the CRs penetrates within the bubble is
$\Delta d \sim \sqrt{2D(E,B)\Delta t_\text{b}} - \Delta R$, where 
$\Delta R\sim 0.5 R_b$ is the thickness of the shell of AGN wind shocked material
(region ii in Figure~\ref{fig:out-shock}).
For a background magnetic field of $B= 5$ G
and $R_\text{b} = 96$ AU for instance (corresponding to the emission model M2, see Table 1),
the penetration depth $\Delta d$ of CR protons with energies
between $1.22$ GeV and $300$ GeV,
is in the range of 
$0.4 R_\text{b} \lesssim \Delta d \lesssim 1.8 R_\text{b}$.
In this case the accelerated CR protons occupy 
almost the whole volume of the bubble.

We parameterise the energy distribution of the CR population within the bubble
as a power-law (P-L) with exponential cut-off of the form
\begin{equation}
    \label{J_p}
    J_\text{p}(E_\text{p}) = 
    A \left(\frac{E_\text{p}}{E_{0}}\right)^{-q} \exp \left\{ -\frac{E_\text{p}}{E_\text{max}} \right\}\,\, \text{erg}^{-1}\text{cm}^{-3},
\end{equation}
where $E_\text{p}$ is the energy of the CR protons (rest mass plus kinetic energy),
$E_{0} = 1.22$ GeV (the threshold energy for $\pi^0$ mesons), $E_\text{max}$ is
the maximum energy of the accelerated protons.
The normalisation constant $A$ is obtained through the condition
\begin{equation}
    \label{normJp}
    0.1 E_\text{w} =\int_V dV \int_{E_\text{min}}^{\infty} dE \,E J_\text{p}(E)\,.
\end{equation}
In this condition, we fix the total energy of the CR population to be 
one tenth of the kinetic wind energy that impinges the surface of the shock
formed by the AGN wind and the outflow bubble during its expansion from 
$R_0$ until $R_\text{b}$ (see equation~\ref{E_w}). 
This 10\%  efficiency is motivated by the energy fraction 
of galactic supernovae needed to explain the galactic CR density 
\citep{2011hea..book.....L}, as well as results of
numerical simulations of particle acceleration by DSA 
(see e. g. \citealt{2014ApJ...783...91C}).

The volume integral in the RHS of equation (\ref{normJp}) is simplified 
assuming that the distribution $J_\text{p}$ is uniform along the volume $V$ of CR emission.
With this consideration, the normalisation constant of the CR distribution (\ref{J_p}) is given
by
\begin{equation}
\label{Anorm}
    A = \frac{0.1 E_\text{w}}{E_0^q V I_\text{p}} \,\,\,\,\mbox{erg}^{-1}\mbox{cm}^{-3},
\end{equation}
where $I_\text{p}(E_0,E_\text{max},q) = \int_{E_0}^{\infty}dE E^{1-q}\exp\{-E/E_\text{max}\}$.

The observed flux of $\gamma$-rays due to the decay of neutral pions produced
by the CR population (\ref{J_p}) in the outflow bubble of OJ 287 is calculated as
\begin{equation}
    \label{pi0flux}
    \nu F_{\nu,\pi^0} = \frac{V}{4\pi D_L^2}
    E'^2 \Phi_\gamma(E') \exp\{-\tau_{\gamma\gamma} (E')\},
\end{equation}
where $E' = h\nu (z+1)$, being $z=$0.306 the redshift of the source, $D_\text{L}= 1602$ Mpc the luminosity
distance, $V$ the volume occupied by the CR injected within the outflow bubble, and $\tau_{\gamma\gamma}$
is the optical depth of photon-photon annihilation within the source given by equation (\ref{tau_gg}).
In equation (\ref{pi0flux}), $\Phi_\gamma$ is the $\gamma$-ray production rate 
(photons per unit energy, per unit time, per unit volume)
in units of erg$^{-1}$ s$^{-1}$ cm$^{-3}$.
For $\gamma$-rays produced by CRs with energies $>100$ GeV
we calculate the function $\Phi_\gamma$ employing 
the parametrisation derived by \cite{2006PhRvD..74c4018K}: 
\begin{equation}
\label{Phi_h}
\Phi_{\gamma,\text{h}}(E')=
cn_\text{b}\int_{E_\gamma}^{\infty} \sigma_\text{pp}(E_\text{p})J_\text{p}(E_\text{p}) 
F_\gamma\left(E'/E_\text{p},E_\text{p}\right)\frac{dE_\text{p}}{E_\text{p}},
\end{equation}
where $n_\text{b}$ is the gas number density of the thermal ions within the outflow bubble (see equation 
\ref{rho_b}),
$\sigma_\text{pp}$ is the total cross section for 
$p$-$p$ interactions given \cite{2006PhRvD..74c4018K} (their equation 79), 
$J_\text{p}$ is the energy density distribution of CR protons defined in equation (\ref{J_p}),
and the function $F_\gamma (E'/E_\text{p}, E_\text{p})$ is defined in equation (58) of \cite{2006PhRvD..74c4018K}.

For $\gamma$-rays produced by CRs with energies $\leq 100$ GeV, we calculate the 
$\gamma$-ray production rate as: 
\begin{equation}
\label{Phi_l}
    \Phi_{\gamma,\text{l}} (E') = \frac{2 c \tilde{n}_\gamma n_\text{b}}{\tilde{K}_\text{pp}}
    \int_{E_{\text{min}\pi^0}}^{\infty} \frac{\sigma_\text{pp}(y)J_\text{p}(y)}{\sqrt{E_\pi^2 - m_\text{p}^2 c^4}} dE_\pi,
\end{equation}
which is a modified version of the $\delta$-functional approach of
\citep{2000A&A...362..937A} suggested in \citep{2006PhRvD..74c4018K}.
In equation (\ref{Phi_l}), $E_{\text{min}\pi^0}=
E' + m_{\pi 0}^2 c^4/(4E')$,
being $m_{\pi 0}$ the mass of the neutral pion, 
and $y \equiv m_\text{p} c^2 + E_\pi/\tilde{K}_\text{pp}$.
Following \citep{2006PhRvD..74c4018K},
$\tilde{K}_\text{pp}=0.17$ is taken as a fixed parameter
(which agrees quite well with numerical Monte Carlo calculations at 
energies $\sim$1 GeV),
and $\tilde{n}_\gamma$ 
(interpreted as the multiplicity of neutral pion production)
is obtained by requiring the functions $\Phi_{\gamma,\text{l}}$ and $\Phi_{\gamma,\text{h}}$
to match at $E' = 0.1$TeV:
\begin{align}
\nonumber    &\tilde{n}_\gamma (q,E_\text{max}) = 
    \frac{\tilde{K}_\text{pp}}{2}\,\,\times\\
\nonumber    &\int_{0.1\text{TeV}}^{\infty}  \sigma_\text{pp}(E_\text{p})J_\text{p}(E_\text{p}, q, E_\text{max} )
    F_{\gamma}(0.1\text{TeV}/E_\text{p}, E_\text{p})\frac{dE_\text{p}}{E_\text{p}}\,\, \times \\
    &\left[\int_{0.1\text{TeV} +m_{\pi0}^2c^4/0.4\text{TeV}}^{\infty}
    \frac{\sigma_\text{pp}(y)J_\text{p}(y)}{\sqrt{E_\pi^2 - m_{\pi0}^2 c^4}}dE_\pi\right]^{-1}.
\end{align}

To calculate the  synchrotron 
and IC emission produced by the secondary $e^{\pm}$ pairs (see Eq. \ref{pipmpairs}), we model
the energy distribution $N_\text{e}(E_\text{e})$ of these leptons (in units of erg$^{-1}$ cm$^{-3}$) 
as a stationary solution of the transport equation \citep[e.g.,][]{1964ocr..book.....G}
for the population of $e^{\pm}$ pairs within the $p$-$p$ emission region:
\begin{equation}
\label{N_e}
    N_\text{e}(E_\text{e}) =
    \left|P_\text{e}\right|^{-1}
    \int_{E_\text{e}}^{\infty} dE_\text{e}' \Phi_\text{e}(E_\text{e}').
\end{equation}
The factor $P_\text{e}$ in equation (\ref{N_e}) is the total
rate of
$e^{\pm}$ energy losses:
\begin{equation}
\label{Pe}
P_\text{e}  = P_\text{syn} + P_\text{IC} + P_\text{br} + P_\text{Co},
\end{equation}
where we consider losses due to synchrotron radiation,
IC scattering,
relativistic bremsstrahlung, and
Coulomb collisions, respectively 
(the expresions for these cooling terms can
be found in Appendix~\ref{appendix:electronCooling}).
The function $\Phi_\text{e}$ within the integral of equation (\ref{N_e}) is the
 $e^{\pm}$ production rate
 in units of erg$^{-1}$ s$^{-1}$ cm$^{-3}$
 (particles per unit energy, per unit time, per unit volume).
For leptons produced by 
CR protons with energies $>0.1$ TeV,
we calculate the function $\Phi_\text{e}$ employing the parametrisation  
derived by \cite{2006PhRvD..74c4018K}:
\begin{equation}
\label{Phi_eh}
    \Phi_\text{e,h}(E_\text{e}) = 
    cn_\text{b}\int_{E_\text{e}}^{\infty} \sigma_\text{pp}(E_\text{p})J_\text{p}(E_\text{p}) 
F_\text{e}\left(E_\text{e}/E_\text{p},E_\text{p}\right)\frac{dE_\text{p}}{E_\text{p}},
\end{equation}
and for leptons produced by CR protons with energies $\leq 0.1$ TeV we employ 
the $\delta$-functional approach (see \citealt{2006PhRvD..74c4018K}):
\begin{equation}
\label{Phi_el}
       \Phi_\text{e,l} (E_\text{e})= 
    \frac{2c n_\text{b} \Tilde{n}_\text{e}}{\Tilde{K}_\text{pp}}
     \int_{E_{\text{min}\pm}}^{\infty}
    \sigma_\text{pp}\left(y\right)
    J_\text{p}\left(y\right)
    f_\text{e}(E_\text{e}/E_\pi) \frac{dE_{\pi}}{E_{\pi}},
\end{equation}
similarly as done in the calculations of  \cite{2016MNRAS.460...44P}.
In equation (\ref{Phi_el}), $E_{\text{min}\pm} = 
E_\text{e} + (m_{\pi\pm} c^2)^2/(4E_\text{e})$, where
$m_{\pi\pm}$ is the mass of charged pions.
The functions $F_\text{e}(E_\text{e}/E_\text{p},E_\text{p})$ and $f(E_\text{e}/E_\pi)$ in equations (\ref{Phi_eh}) and (\ref{Phi_el}) 
are defined by equations (62) and (36) of \cite{2006PhRvD..74c4018K}, respectively.
Similarly as in the case case for $\gamma$-ray production,
we set $\tilde{K}_\text{pp}=0.17$ and the factor $\tilde{n}_\text{e}$ in
equation (\ref{Phi_el}) is obtained from  the condition 
$\Phi_\text{e,h}(E_\text{e} = 0.1$ TeV$) = \Phi_\text{e,l}(E_\text{e} = 0.1$ TeV).

Once the stationary distribution (\ref{N_e}) is computed, we apply it to calculate the 
synchrotron and IC fluxes following the usual prescriptions (e.g., \citealt{1970RvMP...42..237B},
\citealt{2010A&A...519A.109R}).
The flux of synchrotron radiation at the Earth is then calculated as:
\begin{equation}
    \nu F_{\nu, \text{syn}} = \frac{V}{4\pi D_\text{L}^2}E'\int_{E_\text{min,e}}^{E_\text{max,e}}
    dE_e N_e(E_e) \langle P_\text{syn}\rangle_{\alpha}
\label{Fsyn}
\end{equation}
where 
$\langle P_\text{syn}\rangle_{\alpha}=\frac{1}{2}\int_0^\pi \sin\alpha P_\text{syn}d\alpha$  is the 
synchrotron emission power, averaged over the pitch angle $\alpha$, 
\begin{equation}
    P_{\text{syn}} (E_e, E', \alpha, B) = \frac{\sqrt{3}e^3 B \sin\alpha}{h m_e c^2} \frac{E'}{E_c}
    \int_{E'/E_c}^{\infty} K_{5/3}(x)dx,
    \label{P_syn}
\end{equation}
$K_{5/3}(x)$ is the modified Bessel function of order 5/3, and
\begin{equation}
        \label{E_c}
    E_c = \frac{3}{4\pi}\frac{ehB \sin\alpha}{m_e c} \left(\frac{E_e}{m_e c^2} \right)^2.
\end{equation}
In equations (\ref{P_syn}) and (\ref{E_c}), $B$ is the magnetic field in the hadronic emission region (region iii in Figure \ref{fig:out-shock}b)
which we assume to have the same value as in the acceleration region
(region ii in Figure \ref{fig:out-shock}b).

For the emission due to IC scattering of the secondary electron-positron pairs, we consider as
seed photons the thermal radiation of the outflow bubble.
Thus, we calculate the observed flux of IC emission as
\begin{align}
\label{ICflux}
\nonumber    \nu F_{\nu,\text{IC}} = 
    &\frac{V}{4\pi D_\text{L}^2}  \exp\{-\tau_{\gamma\gamma} (E')\} \,\,\times \\
    & E'^2 \int_{E_\text{min,e}}^{E_\text{max,e}} dE_\text{e} N_\text{e}(E_\text{e})
    \int_{\epsilon_\text{min}}^{\epsilon_\text{max}} d\epsilon P_\text{IC}(E',E_\text{e},\epsilon),
\end{align}
with
\begin{equation}
\label{PIC}
    P_{\text{IC}} = \frac{3\sigma_\text{T}  m_\text{e}^2 c^5}{4E_\text{e}^2}\frac{n_{ph}(\epsilon, T_\text{b})}{\epsilon} 
    F(E',E_\text{e},\epsilon).
\end{equation}
In equation (\ref{ICflux}), $N_\text{e}(E_\text{e})$ is the energy distribution of electron-positron pairs
calculated with equation (\ref{N_e}). In equation (\ref{PIC}), $n_{ph}(\epsilon, T_\text{b})$ is the
photon field generated by the thermal radiation  within the outflow bubble of temperature $T_\text{b}$ given by equation (\ref{n_ph}), and
\begin{align}
    &F(E',E_\text{e},\epsilon) =
    2q\ln q + (1+2q)(1-q) + \frac{1}{2}(1-q)\frac{(\Gamma  q)^2}{1+\Gamma q},\\
\nonumber    &q  = \frac{E'}{\Gamma(E_\text{e} - E')},  \\
\nonumber    &\Gamma  = \frac{4\epsilon E_\text{e}}{m_\text{e}^2 c^4},
\end{align}
where the energy of the scattered photons is in the range
$\epsilon \leq E' \leq E_\text{e}\Gamma/(1+\Gamma)$.
We note that to calculate the fluxes due to pion decay, synchrotron 
and IC scattering of secondary $e^{\pm}$ with equations
(\ref{pi0flux}), (\ref{Fsyn}), and (\ref{ICflux}),
it is not needed to explicitly specify
the volume $V$ of the hadronic emission region within the outflow bubble.
This is because, according to the normalisation condition of equation (\ref{normJp}),
the energy distributions of CR protons and electron-positron pairs are
$J_\text{p}\propto V^{-1}$ and $N_\text{e}\propto V^{-1}$,
which cancel the volume factor in equations (\ref{pi0flux}), (\ref{Fsyn}), and (\ref{ICflux}).

In the next Section, we apply the non-thermal and thermal emission processes described in
this and the previous section, to model the observed MW SED corresponding to the November 2015 flare of OJ 287.

\section{SED models for the 2015 major flare of the blazar OJ 287}
\label{sec:SED_models}

Following its historical $\sim$12 year optical flares,  OJ 287 displayed a major optical excess in November 2015
 in agreement with the prediction of the SMBH binary model
\citep{2016ApJ...819L..37V}.

In the X-ray and $\gamma$-ray bands, flare activity was also reported.
\cite{2018MNRAS.473.1145K} carried out a MW analysis of the LCs during and after the November 2015 flare finding
significant activity most prominently in the NIR, optical, UV and X-ray bands, associated with significant change in the polarisation angle (PA) and polarisation degree
\citep[PD; see also][]{2019AJ....157...95G}.
 \cite{2018MNRAS.473.1145K} extracted the MW SED of the flare and interpreted the X-ray and $\gamma$-ray  components in terms of leptonic jet emission
(they find X-rays consistent with SSC whereas $\gamma$-ray data are better explained with EC),
and the optical component in terms of  multi-temperature disc emission.

Here, we present an alternative model for the SED extracted by \cite{2018MNRAS.473.1145K}.
Motivated by the fact that the LCs in the  X-ray and $\gamma$-ray bands display flaring
simultaneously with the optical excess, we interpret this flare state 
in terms of the BH-disc impact scenario 
described in Sections \ref{sec:bh-disc}-\ref{sec:non-thermal}.
To do this, we first determine the properties of the outflow bubble
($R_\text{b}, n_\text{b}, T_\text{b}$, and $n_\text{ph}$)
when the outburst occurs. This is done
by matching the thermal bremsstrahlung emission of the bubble
with the optical flare data using $v_\text{r}$, $n_\text{d}$, and $f_\text{R}$ 
as free parameters (see Section~\ref{sec:scenario}).  
Then, we calculate the non-thermal emission of the outflow bubble
with the hadronic emission model described in Section \ref{subsec:ppemit}.

\begin{figure}
    \includegraphics[width=8.4cm]{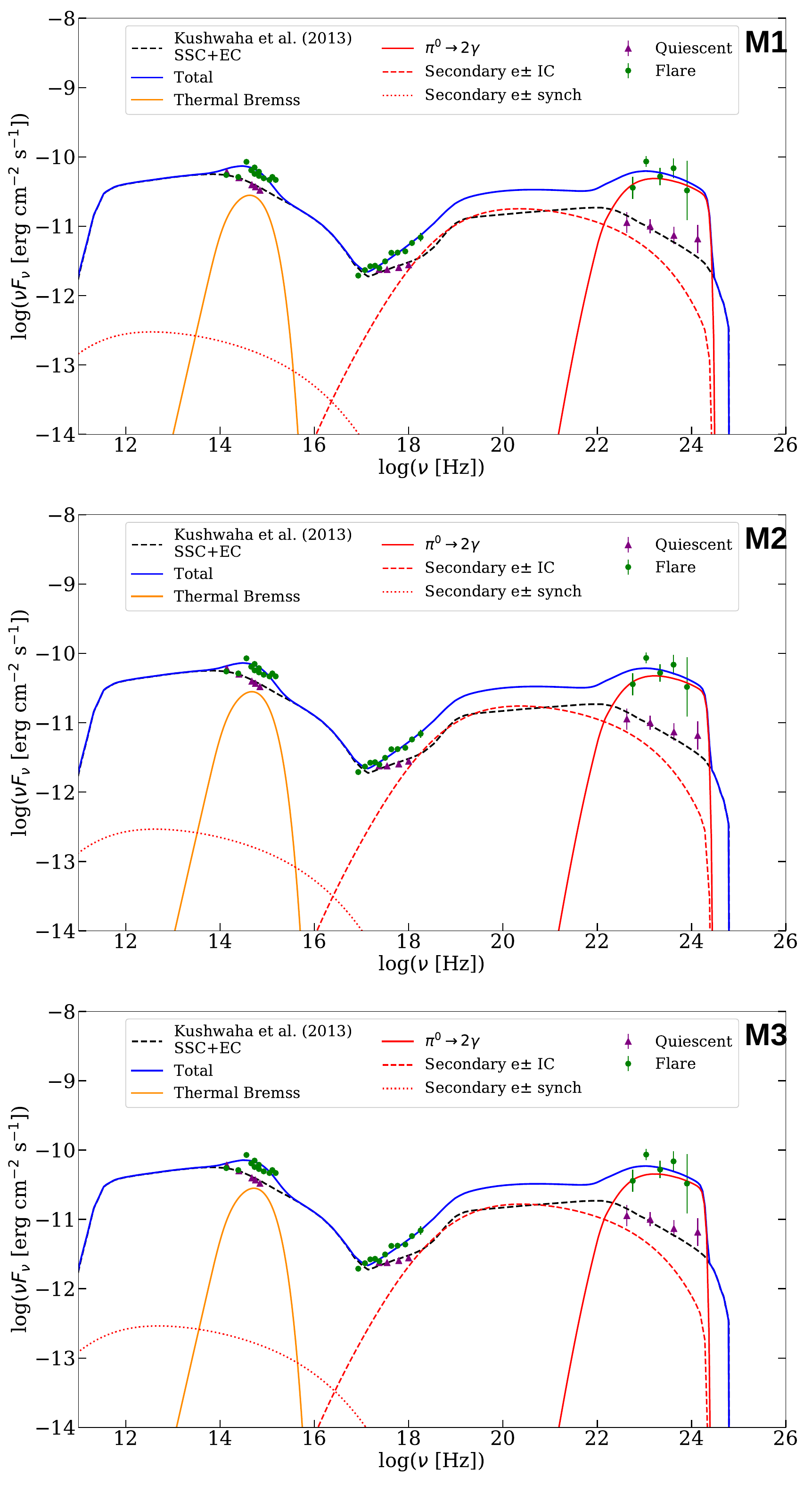}
    \caption{
    Flare and quiescent SEDs of the blazar OJ 287. The flare data 
    (green points, adapted from \citealt{2018MNRAS.473.1145K})
    corresponds to the period MJD: 57359-57363
    (simultaneous to the November 2015 optical major flare).
    The data of the quiescent state
    (magenta points, adapted from \citealt{2013MNRAS.433.2380K})
    corresponds the period MJD: 55152-55184.
    The over-plotted curves are emission models for the quiescent 
    (black dashed curve, adapted from \citealt{2013MNRAS.433.2380K})
    and flare state (blue solid curve). 
    The red and orange curves are the components of the flare
    thermal+hadronic emission model derived in this paper (see the text).
    The three panels display the same data points and the same quiescent emission model,
    whereas a different flare emission profile is displayed in each panel.
    The model parameters of the flare emission profiles (labelled as M$i$) are listed in Table~1.
}
    \label{fig:SED_Ms}
\end{figure}

To define the magnetic field $B$ in the acceleration region,
some features of the broadband flare data together with the model described in
Section~\ref{subsec:ppemit} offer the following constrains.
(i) We note that the magnetic field B in the acceleration region must be high enough to
accelerate CR protons up to an energy $E_\text{max}$ able to produce $\gamma$-ray photons
of at least $\sim$5 GeV, the highest energy of the $\gamma$-ray data.
This is a lower limit for $E_\text{max}$, since $\gamma$-ray photons with energies much higher
than $\sim$5  GeV could be produced but not seen due to photon-photon annihilation (see Section~\ref{subsection:photon-photon}).
(ii) The magnetic field must be able to cool the secondary e$\pm$ pairs enough
(by synchrotron losses) for these leptons to not overproduce 
X-ray photons by IC scattering.
(iii) At the same time, the magnetic field should be low enough to
produce a negligible flux of synchrotron radiation at optical energy bands.
This last  condition is imposed by the initially low PD observed in the optical outburst.
Thus, once we obtain the outflow bubble properties ($R_\text{b}, T_\text{b}, n_\text{b}, n_\text{ph}$),
we use $E_\text{max}$ as a free parameter to derive the
magnetic field $B$ through the balance equation~(\ref{rate_balance})
(for acceleration and cooling rates of CR protons).
With this procedure, we seek for parameter configurations
implying broadband SEDs fulfilling the 
conditions listed above and at the same time requiring an AGN wind power as low as possible.

\begin{table}
\label{para}
\caption{Free and derived  parameters for the SEDs profiles of the models M1, M2 and M3 displayed
in Figure~\ref{fig:SED_Ms}. (See more details in the text.)} 
 \centering
 \begin{tabular}{|l| l l l l |}
\hline\hline
 \multicolumn{2}{r}{}      & \multicolumn{1}{c}{M$_1$} & \multicolumn{1}{c}{M$_2$} &
 \multicolumn{1}{c}{M$_3$}  \\
\cline{1-5}
\multirow{11}{*}{\rotatebox[origin=c]{90}{Free}}
\rule{0pt}{15pt}  & $n_{\text{d}}\left[10^{14}\mbox{cm}^{-3}\right]$    & 1.00        & 1.00     &  1.00      \\
\rule{0pt}{15pt} & $(v_\text{r}/c)\times10$   &  1.54       & 1.62      & 1.70    \\
\rule{0pt}{15pt} & $f_\text{R}$               &  0.85       &  0.85     &  0.85    \\
\rule{0pt}{15pt}  & $n_{\text{w}}\left[10^6\mbox{cm}^{-3}\right]$    & 6.50        & 6.50     &  6.50       \\
\rule{0pt}{15pt} & $T_\text{w} \left[10^9\mbox{K}\right]$      &  1.00       & 1.00      & 1.00   \\
\rule{0pt}{15pt} & $(V_\text{w}/c)\times10$                    &  2.20       & 2.50      & 2.80    \\
\rule{0pt}{15pt}  & $q$                                        &  2.20       &  2.20     &  2.20     \\
\rule{0pt}{15pt}  & $E_\text{max} \left[\text{TeV}\right]$     &  0.30       &  0.30     &  0.30     \\
\cline{1-5}
\multirow{20}{*}{\rotatebox[origin=b]{90}{Derived}}
 \rule{0pt}{15pt} & $T_\text{0}\left[10^6 \mbox{K}\right]$ & 1.05       & 1.08      & 1.10    \\
\rule{0pt}{15pt}  & $R_\text{0}$ [AU]                       & 106.57     & 96.30      &  87.46    \\
\rule{0pt}{15pt}  & $\xi = R_\text{b}/R_\text{0}$         & 42.44    & 39.05     &  36.08  \\
\rule{0pt}{15pt}  & $\Delta t_\text{b}$ [yr]                      & 1.88     & 1.43      &   1.09 \\
\rule{0pt}{15pt}  & $n_\text{b}\left[10^{10}\mbox{cm}^{-3}\right]$    & 0.92     & 1.18      &  1.49    \\
\rule{0pt}{15pt}  & $T_\text{b}\left[10^4 \text{K}\right]$         & 2.47     & 2.76      &  3.06     \\
\rule{0pt}{15pt}  & $\frac{L_\text{w}}{\dot{M}_1 c^2}\times10$    & 2.46     & 3.61     &  5.07    \\
\rule{0pt}{15pt} & $ B$ [G]                               &  4.20      & 5.08      & 6.20      \\
\rule{0pt}{15pt}  & $\mathcal{M}_{\text{sA}}$            & 10.63    & 11.85     &  12.97  \\
\rule{0pt}{15pt}  & $E_{\text{CR}}\left[10^{52}\mbox{erg}\right] $  & 4.75    & 3.59     &  2.70     \\
 \cline{1-5}
 \end{tabular}
 \end{table}

The green data points in the plots of Figure \ref{fig:SED_Ms} are the flare MW SED extracted from the LCs data corresponding to MJD: 57359-57363 (see \citealt{2018MNRAS.473.1145K}).
The data points in magenta represent the SED of what we consider as the pre-burst, quiescent state,
for which we take the SED data extracted by \cite{2013MNRAS.433.2380K} corresponding to the 2009 broadband LCs of the source (their ``state 3'' data, when no significant variability was displayed).
The flare data (green points) 
are from: 
\textit{Fermi}-LAT \citep{2009ApJ...697.1071A} at $\gamma$-ray energies,
\textit{Swift}-XRT \citep{2005SSRv..120..165B} at X-rays, and
Swift-UVOT + 11 ground-based observatories 
\citep{2017MNRAS.465.4423G,2018MNRAS.473.1145K}. 
Similarly for the quiescent state data (magenta points).

In Figure~\ref{fig:SED_Ms}, we overplot different SED emission profiles that result from the
thermal+hadronic emission model and the parameters associated to each model are listed in Table~1.
The blue curve is the calculated total emission that results from the quiescent plus the flare components.
The quiescent state is a SSC + EC jet emission model (black, dashed curve), which here we adapt
from  \cite{2013MNRAS.433.2380K}.
The orange solid curve is the outburst thermal bremsstrahlung emission (described in Section 2), 
whereas the red curves represent the flare emission of hadronic origin (see Section \ref{subsec:ppemit}).  
The red solid curve corresponds to $\pi^0$ decay emission. The red
dotted and dashed curves correspond to synchrotron and IC fluxes of the secondary $e^\pm$ pairs, respectively.
In all the models derived here, the thermal bremsstrahlung radiation is more
intense, by two orders of magnitude or more, than the synchrotron radiation
generated by the $e^{\pm}$ pairs. Because of this, we neglect the SSC contribution,
and thus we employ the thermal bremsstrahlung photon field given by equation (\ref{n_ph})
as the dominant source of seed photons for IC scattering.

The models M1, M2, and M3 share the same assumed free parameters with the exception of
$v_\text{r}$ (the velocity of the disc material relative to the secondary SMBH at the impact event)
and $V_\text{w}$ (the AGN wind velocity).
We note that the emission models are most sensitive to the variations of $v_\text{r}$.
For larger $v_\text{r}$, a larger AGN wind power is required to match the HE data.
Adopting different values of $v_\text{r}$ and $V_\text{w}$ we generate emission profiles
consistent with the broadband data which result in outflow bubbles with noticeable different properties
(see the derived parameters in the second section of Table~1).

In all the models presented in Figure~\ref{fig:SED_Ms}, the synchrotron emission of the secondary $e^\pm$
pairs does not contribute substantially to any spectral region of the flare data.
This is particularly consistent with the low PD initially observed in the optical flare.
We also see that the X-ray and $\gamma$-ray spectral components are consistent with a CR population
(within the outflow bubble) of P-L index $q= 2.2$ and maximum energy $E_\text{max}=0.3$ TeV.

According to the displayed emission models, to reproduce the HE flare SEDs,
an AGN wind representing $\lesssim$50\% of the primary SMBH accretion power
is required (according to the definition in equation~\ref{eta_w}). 
Considering energy efficiency, the model M1 appears to be 
the most favoured.

\section{Summary and discussion}
\label{sec: discussion} 

The BL Lac blazar OJ 287 displayed a major optical outburst in November 2015, 
in agreement with its well known  $\sim$12 yr optical periodicity.
Simultaneous flaring was also reported in the X-ray and $\gamma$-ray bands, 
with hardening in the spectral index compared with previous states seen in 
this source \citep{2018MNRAS.473.1145K}.
In the present work, we show that a hadronic emission component compatible with 
the presence of a SMBH binary system in the core of OJ 287
reproduces the  X-ray and $\gamma$-ray data self-consistently
with a thermal component constrained by the optical outburst.

The one-zone thermal+hadronic emission model presented here is based on the following considerations.
\begin{itemize}
\item 
The secondary SMBH impacts the accretion disc of the primary one, 
generating a bipolar outflow 
\citep{1998ApJ...507..131I,2016MNRAS.457.1145P}.
The outflow that emerges the accretion disc in the direction of
the observer is the dominant source of the observed emission.
We model this outflow following \cite{1996ApJ...460..207L}, 
considering it as a spherical bubble 
that grows at the speed of sound of the internal gas.

\item 
CR protons are accelerated in the forward shock formed by the expanding  
outflow bubble as it collides with the local AGN wind (see Section~\ref{sec:non-thermal}). 

\item 
By the time of the optical outburst 
a population of CR protons has been injected within the outflow bubble (see Figure~\ref{fig:out-shock}b).
This CR population has a total energy
representing one tenth of the AGN wind kinetic energy that impinged on the outflow bubble
during its expansion.
We then calculate the $\pi^0$ decay emission as well as synchrotron and
IC scattering of secondary $e^\pm$ pairs, where the neutral pions and secondary
leptons are generated out of $p$-$p$ interactions 
\citep{2000A&A...362..937A,2006PhRvD..74c4018K}. 
The dominant seed photon field for IC scattering is the thermal bremsstrahlung 
radiation of the outflow bubble (see Section \ref{sec:non-thermal}).
\end{itemize}

We present different emission profiles which can explain the observed flare data.
The inferred magnetic field in the acceleration region of these emission models
is in the range of $\sim 4 - 6$ G.
The CR population responsible for the X-ray and $\gamma$-ray components
is consistent with a P-L index $q\sim 2.2$ and a cut-off energy of $\sim 0.3$ TeV.
In all the derived models, the mechanical luminosity of the AGN wind 
represents $\lesssim 50\%$ of the mass accretion power of the primary SMBH (see Table~1).

To interpret the emission data with the one-zone hadronic 
emission model discussed here, we assumed values for the size of the
corona region as well as for the parameters of the AGN wind which appear to be
reasonable and according to previous studies of coronae and wind of AGNs (see Section~\ref{subsec:ppemit}).
However, more realistic models for the corona and the AGN wind (which are beyond the scope
of this paper) could, perhaps, modify the results obtained here. 
For instance, the maximum energy $E_\text{max}$ of accelerated  protons by DSA depends on the
velocity and magnetic field of the wind impinging on the outflow bubble.
Also, due to the precessing nature of the secondary SMBH orbit, 
the BH-disc impacts are expected to occur at different radii from the primary SMBH 
\citep[see e. g.][]{2018ApJ...866...11D} and DSA  would not be  efficient for BH-disc
impacts occurring closer (or inside) to the primary SMBH AGN corona 
(see Section~\ref{subsec:shock}).
Such study of whether or not closer impacts produce observable 
HE emission, we leave for a future work.

The acceleration, propagation and emission of CRs within the source
depend on the diffusion of these relativistic particles. For
simplicity, here we adopted a spatially uniform, Kolmogorov-like
diffusion coefficient (see Section~\ref{subsec:ppemit}).
Nevertheless, more elaborate scenarios for CR diffusion could
lead to more efficient particle acceleration, implying higher maximum
energies for CR protons and detectable fluxes of HE neutrinos\footnote{
OJ 287 is in declination favourable for detection with IceCube
and estimated as a potential source of HE neutrinos
based on a different emission scenario than the one discussed here
\citep{2019MNRAS.489.4347O}.
}.
For example, enhanced turbulence is expected to develop behind the
forward shock \citep{2007ApJ...663L..41G,2014MNRAS.439.3490M}, increasing
the CR confinement and reducing the acceleration time. Additional
increase in the acceleration efficiency can take place if the magnetic
field in the wind is amplified ahead of the shock, for instance due to
the vorticity generated by the interaction between the CR pressure and
density fluctuations in the supersonic wind  \citep{2009ApJ...707.1541B,2016MNRAS.458.1645D}.

If the non-thermal emission scenario discussed here is indeed correct,
future simultaneous broadband outbursts (if detected) 
will further constrain the properties of the claimed SMBH binary in OJ 287.
Also, the data of future outbursts may be applied to test shock acceleration models 
as well as more realistic 
multidimensional models of AGN-winds and coronae.

\section*{Acknowledgements}
We are thankful to the anonymous reviewer for providing valuable critics which
improved the quality of this work. 
We acknowledge support from the Brazilian agencies FAPESP (JCRR's grant: 2017/12188-5, PK's grant: 2015/13933-0,  and EMGDP's grant: 2013/10559-5), CNPq (EMGDP's grant: 306598/2009-4 ), 
and PK's ARIES Aryabhatta Fellowship (AO/A-PDF/770).

\section*{Data Availability}

The datasets employed in this research were derived in the papers by 
\cite{2013MNRAS.433.2380K}, and
\cite{2018MNRAS.473.1145K} from sources in the public domain: 
High Energy Astrophysics Science Archive Research Centre (HEASARC;
\url{https://fermi.gsfc.nasa.gov/ssc/})
maintained by the National Aeronautics and Space Administration (NASA), 
SMARTS Optical/IR Observations of Fermi Blazars
(\url{http://www.astro.yale.edu/smarts/glast/home.php}), and 
Steward Observatory spectropolarimetric monitoring project
(\url{http://james.as.arizona.edu/~psmith/Fermi/}).
The data underlying this article will be shared on request.



\bibliographystyle{mnras}
\bibliography{refs} 




\appendix

\section{The energy stored by the BH-disc collision}
\label{appendix:E_outflow}
\begin{figure}
	\includegraphics[width=8.5cm]{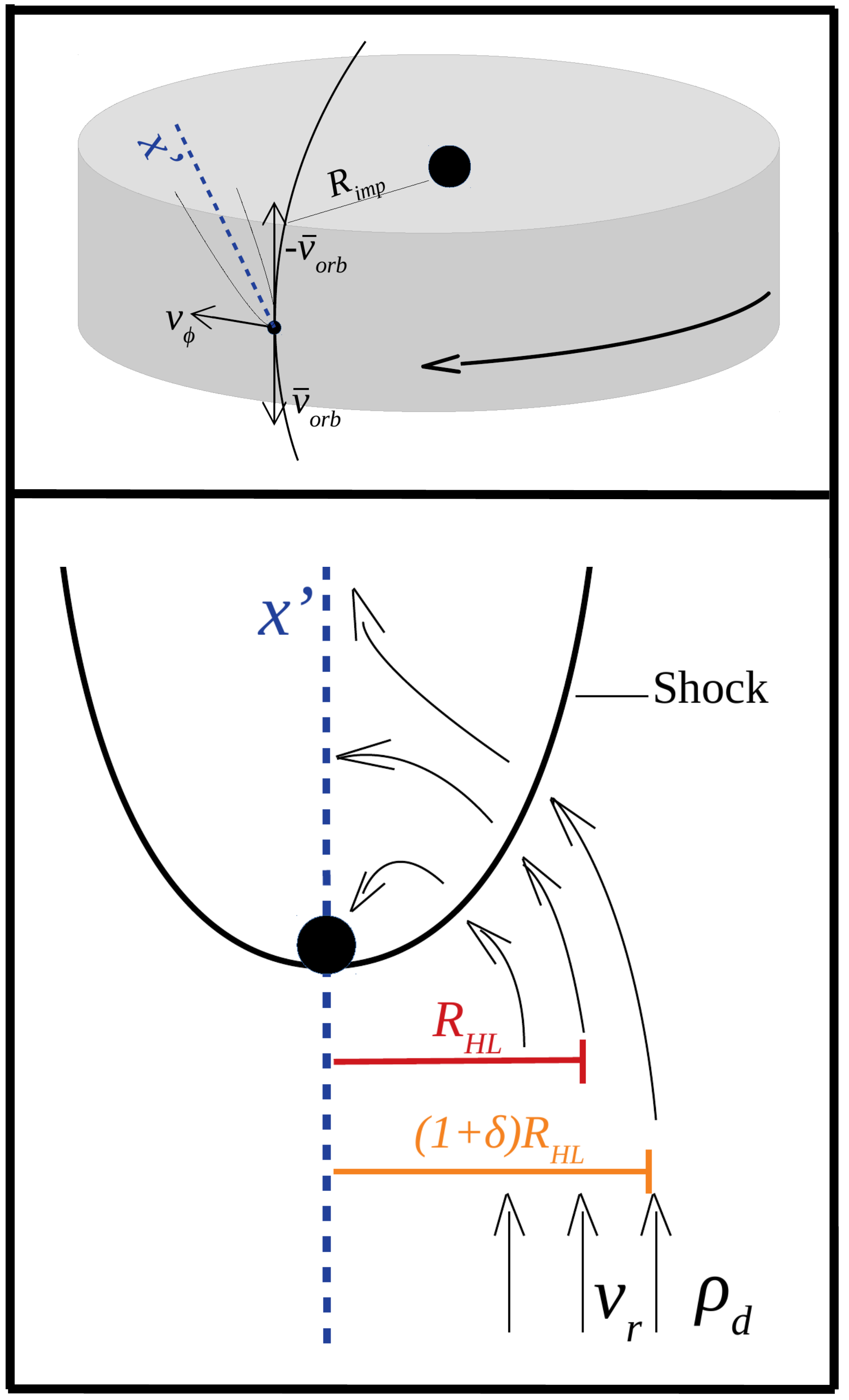}
    \caption{
    Upper:
    Schematic illustration of Bondi-Hoyle-Lyttleton  accretion onto the secondary BH (the small black circle)
    while traversing the accretion disc of the primary (the big black circle, see the text).
    The grey cylinder represents the central volume of the accretion disc
    within the radius $R_\text{imp}$ (the radius at which the secondary BH impacts the disc).
    The arrows represent the velocity of the travelling BH ($\vec{v}_\text{orb}$) and the toroidal
    component of accretion disc fluid velocity ($\vec{v}_{\phi}$).
    Lower: Schematic view in the co-moving frame of the travelling BH.
    In this frame the disc material of density $\rho_d$ impinges the secondary BH with 
    velocity $v_\text{r} = |\vec{v}_\phi-\vec{v}_\text{orb}|$.
    The red bar represents the Bondi-Hoyle-Lyttleton  radius  $R_\text{HL}$(see equation \ref{R_HL}).
    Particularly, the material impinging at radii between $R_\text{HL}$ and $(1+\delta)R_\text{HL}$
    is not accreted onto the travelling BH.}
    \label{fig:delta_RHL}
\end{figure}

Here we estimate the energy injected by the secondary BH after its passage through 
the disc of the primary one. To do this, we assume that Bondi-Hoyle-Lyttleton (BHL) 
accretion theory \citep{1939PCPS...35..405H, 2004NewAR..48..843E, 2011MNRAS.417.2899Z}
describes well the accretion of the disc material onto the travelling BH. 
As noted by \cite{1998ApJ...507..131I} and \cite{2016MNRAS.457.1145P}, this turns out to be the case if
the thickness $\Delta H$ of the accretion disc is
$\Delta H >> R_\text{HL}$ at the location of the impact,
where $R_\text{HL}$ the  Bondi-Hoyle-Lyttleton radius of the travelling BH:
\begin{equation}
    R_\text{HL} = \frac{2 G M_2}{v_\text{r}^2},
    \label{AR_HL}
\end{equation}
and $v_\text{r}$ the resultant velocity of the disc material in the co-moving frame of the travelling BH
(see Figure~\ref{fig:delta_RHL}, upper).
Thus, the estimate described in the following,
is more reliable the smaller the ratio $R_\text{HL}/\Delta H$.

A wake of shocked material is formed behind the travelling BH provided that its velocity is
supersonic with respect to the sound speed of the disc fluid, which is the case
for the situation discussed here.
In the co-moving frame of the secondary BH, the upstream flow
impinging at cylindrical radii
$\leq R_\text{HL}$ is eventually accreted onto the gravitational source 
(see Figure~\ref{fig:delta_RHL}, lower panel). 
On the other hand, there is an amount of upstream gas impinging 
at cylindrical radii $>R_\text{HL}$ 
that is deflected by the BH gravity,
is compressed through the shock (downstream), 
and have enough kinetic energy to not fall onto the BH as illustrated 
in Figure \ref{fig:delta_RHL} (lower panel).

We consider $\Delta t = \Delta H /v_\text{orb}$ as the time scale of the impact event,
where $v_\text{orb}$ is the velocity of the secondary BH at the location of the impact.
Additionally, we consider that the  kinetic energy of the upstream flow impinging 
at cylindrical radii between $R_\text{HL}$ and $(1+\delta)R_\text{HL}$
(where $\delta$ is a control dimensionless parameter)
during the time interval $\Delta t$
will eventually drive the outflows that emerge from the disc. 
Due to the gravity of the secondary BH,
the material flowing upstream through the annulus defined by the radii 
$R_\text{HL}$ and $(1+\delta)R_\text{HL}$
converges downstream of the BH
(as illustrated in Figure~\ref{fig:delta_RHL}, lower panel).
Motivated by the numerical simulations of \cite{1998ApJ...507..131I}, 
here we assume that this converging material eventually
split in two outflows that emerge above and bellow the accretion disc.
The morphology of the emerging outflows may be initially highly asymmetric. 
Nevertheless, to proceed analytically we consider the emerging outflows as
expanding spheres with physical properties (such as temperature, gas density, and radius)
equivalent to the average properties of the ``real'' outflows.

The kinetic energy of the upstrean flow impinging 
between $R_\text{HL}$ and $(1+\delta)R_\text{HL}$ can be estimated as
 \begin{equation}
 E_\text{K} = A \Delta t \frac{dE}{dA dt},
 \label{E_K}
 \end{equation}
where $A=\pi(2\delta+\delta^2)R_{\text{HL}}^2$ and
$\frac{dE}{dtdA}=\frac{1}{2}\rho_\text{d} v_{\text{r}}^3$ is
the flux of (the upstream) kinetic energy.
Thus, assuming conservation of energy and that the energy given by
equation~(\ref{E_K}) equally split among the two outflows that emerge,
the energy $E_0$ of each outflow is:
\begin{equation}
E_0 = \frac{\pi}{4}(2\delta + \delta^2)
\left(\frac{\Delta H}{R_{\text{HL}}}\right)
\left(\frac{v_{\text{r}}}{v_{\text{orb}}}\right)
R_{\text{HL}}^3 \rho_\text{d} v^2_{\text{r}},
    \label{AppE01}
\end{equation}
Similarly, assuming conservation of mass, the mass $M_0$ of each outflow is
\begin{equation}
    M_0 = \frac{\pi}{2}
    \left(2\delta + \delta^2\right)
    \left(\frac{\Delta H}{R_{\text{HL}}}\right)
    \left(\frac{v_{\text{r}}}{v_{\text{orb}}}\right)
    R_{\text{HL}}^3 \rho_\text{d}.
    \label{M0flux}
\end{equation}
As described in the main text of this paper, we parametrise the outflows
as spherical blobs that emerge the disc with
initial radius $R_0 = f_\text{R} R_{\text{HL}}$, ($f_{\text{R}}\leq 1$)
and initial gas density $\rho_0 = 7\rho_{\text{d}}$ 
(given by the compression of a strong, radiation dominated shock).
In terms of this parametrisation, the mass of the each blob can be expressed as
\begin{equation}
    M_0 =\frac{4\pi}{3}
    \left( f_{\text{R}}R_{\text{HL}} \right)^3
    7 \rho_\text{d}.
    \label{M0para}
\end{equation}
Combining equations (\ref{M0flux})-(\ref{M0para}), the dimensionless 
control parameter $\delta$ is given by 
\begin{equation}
    \left( 2\delta + \delta^2 \right)
    \left(\frac{\Delta H}{R_{\text{HL}}}\right)
    \left(\frac{v_{\text{r}}}{v_{\text{orb}}}\right)
    =
    \frac{56}{3}f_{\text{R}}^3
    \label{eqdelta},
\end{equation}
and combining ~(\ref{eqdelta}) with equation~(\ref{AppE01}), 
the energy of each blob can be expressed as
\begin{equation}
    E_0= \frac{14\pi}{3}
    \left(
    f_{\text{R}} R_\text{HL}
    \right)^3
    \rho_{\text{d}} v^2_\text{r}.
    \label{AppE02}
\end{equation}

\section{Energy losses of CR protons}
\label{appendix:rates_protons}

The rate of energy losses due to diffusion of CR protons in equation~(\ref{rate_balance}) is calculated as 
\begin{equation}
t_\text{diff}^{-1} = 2 D (E_\text{p}, B) / R_\text{b}^2, 
\end{equation}
where $D$ is the diffusion coefficient defined in equation~(\ref{D}) and 
$R_\text{b}$ is the radius of the outflow bubble.

The rate of energy losses due to proton-proton interactions is computed as:
\begin{equation}
\label{tppf}
    t_\text{pp}^{-1}(E_\text{p},n_0) = 
     K_\text{pp}c n_0\sigma_\text{pp}(E_\text{p}), 
\end{equation}
where $K_\text{pp}\sim0.5$ is the inelasticity factor, $\sigma_\text{pp}$ is the total cross section
for $p$-$p$ interactions taken from \cite{2006PhRvD..74c4018K}, and $n_0$ is the local density
of thermal ions.
We employ equation (\ref{tppf}) to calculate the rate of proton-proton interactions
in the region within the outflow bubble as well as in the shell of the shocked AGN 
wind material.
For the region within the outflow bubble we set $n_0 = n_\text{b}$ (the density of the bubble, see equation~\ref{rho_b}).
Distinctly from the gas of the bubble, the gas in the shell of shocked AGN wind is not
radiation pressure dominated (see equation~\ref{n_sh} and related text).
Thus, in the swept up shell we set $n_0= 4 n_\text{w}$, where $n_w$ is the 
gas number density of the impinging AGN wind. This corresponds to the case of a strong shock
and the specific heat ratio of $\gamma_\text{sh}=5/3$.

The rate of energy losses of protons due to photo-pion production is computed as
\citep[e.g.,][]{2003ApJ...586...79A, 2010A&A...519A.109R}:
\begin{equation}
    t_{\text{p}\gamma}^{-1}(E_\text{p}) = 
    \frac{m_\text{p}^2 c^5}{2 E_\text{p}^2}
    \int_{\frac{\epsilon_\text{th}}{2\gamma_\text{p}}}^{\infty} d\epsilon \frac{n_\text{ph}(\epsilon)}{\epsilon^2}
    \int_{\epsilon_\text{th}}^{2\epsilon\gamma_\text{p}} d\epsilon' \epsilon' 
    K_{\text{p}\gamma}(\epsilon')\sigma_{\text{p}\gamma}(\epsilon'),
\end{equation}
where, $\epsilon_\text{th}=145$ MeV is the photon energy threshold for pion production in the rest-frame of the incident proton. 
For the inelasticity and the 
total cross section of the interaction,
$K_{\text{p}\gamma}$ and
$\sigma_{\text{p}\gamma}$,
we follow the approximation given by 
\cite{2003ApJ...586...79A}.

\section{Electron energy losses}
\label{appendix:electronCooling}
According to equation~(\ref{N_e}), we
calculate the energy distribution of secondary $e^\pm$ pairs 
by considering the energy losses of this particle population due to
synchrotron radiation $P_{\text{syn}}$, 
inverse Compton scattering $P_{\text{IC}}$,
relativistic bremsstrahlung $P_{\text{br}}$, and 
Coulomb collisions $P_{\text{Co}}$.

We calculate the energy losses due to  synchrotron cooling as:
\begin{equation}
    P_{\text{syn}} = \frac{4}{3} \sigma_\text{T} c \frac{B^2}{8\pi}
    \left(
    \frac{E_\text{e}}{m_\text{e} c^2}
    \right)^2,
\end{equation}
with $\sigma_\text{T}=6.652$ cm$^{2}$ the Thomson cross section, $m_\text{e}$ the electron rest mass and $B$ the 
local magnetic field density.

The IC seed photon field considered here (the thermal bremsstrahlung radiation from the outflow bubble; see Section~\ref{sec:scenario})
has cut-off at $\sim$ 1 eV. Thus for secondary electrons with energy $\lesssim 10^{11}$ eV, the IC scattering occurs in the Thompson regime, and their IC energy losses rate is given by: 
\begin{equation}
    P_{\text{IC}} = \frac{4}{3} \sigma_\text{T} c U_{\text{ph}}
    \left(
    \frac{E_\text{e}}{m_\text{e} c^2}
    \right)^2,
\end{equation}
where
\begin{equation}
U_{\text{ph}}=\int_{\epsilon_{\text{min}}}^{\epsilon_{\text{max}}} \epsilon n_{\text{ph}}(\epsilon) d\epsilon,
\end{equation}
is the energy density of the photon field in the emission region of the $e^\pm$ pairs,
and $n_{\text{ph}}(\epsilon)$ is the photon field density generated by the thermal bremsstrahlung radiation of the 
outflow bubble (see equation \ref{n_ph}).

Assuming a fully ionised medium, the cooling of $e^{\pm}$ by relativistic bremsstrahlung is calculated as 
\citep[e.g.,][]{1997ApJ...490..619S}:
\begin{equation}
    P_{\text{br}} = \frac{8 e^6 n_\text{b}}{\hbar m_\text{e}c^2} 
    \left(
    \ln \left\{\gamma_\text{e}\right\} +0.36
    \right)
    (\gamma_\text{e} + 1),
\end{equation}
with $\gamma_\text{e} = E_\text{e}/(m_\text{e}c^2)$.

The energy loss rate due to Coulomb collisions of CR electrons with
background thermal electrons is calculated as \citep[e.g.,][]{1994A&A...286..983M,1997ApJ...490..619S}:
\begin{equation}
    P_{\text{Co}}
    =
    \frac{2 \pi e^4}{m_\text{e} c} \frac{n_\text{b} \lambda_\text{Co}}{\beta_\text{e}}
    \left[
    \psi(x) - \psi'(x)
    \right]
    ,
\end{equation}
where $\lambda_\text{Co}\sim 15$ is the Coulomb logarithm for the parameters of the
problem of this work, $\beta_\text{e} = v_\text{e}/c = \sqrt{1-1/\gamma_\text{e}^2}$ is the velocity of CR electrons in units of the speed of light, and 
\begin{align}
\psi (x) = &   \frac{2}{\sqrt\pi} 
\int_0^{x} y^{1/2}\exp\{-y\} \,dy,\\
\psi'(x) = & \frac{d\psi}{dx}, 
\end{align}
where $x = m_\text{e} v_\text{e}^2/(2kT_\text{e})$, being $T_\text{e}$ the electron temperature inside the bubble given by
equation~(\ref{T_b}).

\bsp	
\label{lastpage}
\end{document}